\documentclass[twocolumn]{aastex62}
\usepackage{aas_macros}
\usepackage{float} \usepackage{amsmath,amssymb}
\usepackage{natbib} 
\usepackage{graphicx} 
\usepackage{color} \usepackage{verbatim}

\newcommand{\figdir}{./} 
\newcommand{\proptosim}{\mathrel{\vcenter{
 \offinterlineskip\halign{\hfil$##$\cr
 \propto\cr\noalign{\kern2pt}\sim\cr\noalign{\kern-2pt}}}}}

\newcommand{\unit}[1]{{\rm\, #1}}

\newcommand{\mean}[1]{\langle #1\rangle}
\newcommand{\method}{L. Wang 2018, in preparation}
\renewcommand{\min}{\mathrm{min}}
\renewcommand{\max}{\mathrm{max}}
\newcommand{\au}{\unit{au}}
\newcommand{\cm}{\unit{cm}}

\newcommand{\g}{\unit{g}}
\newcommand{\G}{\unit{G}}
\newcommand{\K}{\unit{K}} 
\newcommand{\km}{\unit{km}}
\newcommand{\kms}{\unit{km\ s^{-1}}}

\newcommand{\erg}{\unit{erg}}
\newcommand{\eV}{\unit{eV}}
\newcommand{\keV}{\unit{keV}}
\newcommand{\s}{\mathrm{s}}
\newcommand{\yr}{\mathrm{yr}}

\newcommand{\ang}{\ensuremath{\mathrm{\AA}}}

\newcommand{\am}{\mathrm{Am}}
\newcommand{\cs}{c_{\rm s}}

\newcommand{\lya}{\text{Ly}\ensuremath{\alpha~}}

\newcommand{\B}{\mathbf{B}}     
\newcommand{\E}{\mathbf{E}}     
\newcommand{\J}{\mathbf{J}}     
\renewcommand{\v}{\mathbf{v}}   
\newcommand{\kb}{k_\mathrm{B}}
\newcommand{\sigb}{\sigma_\textsc{sb}}
\renewcommand{\d}{\mathrm{d}}
\newcommand{\e}{\mathrm{e}}

\renewcommand{\ion}[2]
{{\rm#1}\;\textsc{#2}}
\newcommand{\p}{\mathrm{p}}     
\renewcommand{\mid}{\mathrm{mid}}     

\newcommand{\acc}{\mathrm{acc}}

\newcommand{\eff}{\mathrm{eff}}
\newcommand{\ah}{\mathrm{ah}}   

\newcommand{\tot}{\mathrm{tot}}
\newcommand{\wind}{\mathrm{wind}}

\newcommand{\wb}{\mathrm{wb}}
\newcommand{\dust}{\mathrm{dust}}
\newcommand{\disk}{\mathrm{disk}}

\newcommand{\euv}{\mathrm{EUV}}

\renewcommand{\O}{\mathrm{O}}     
\newcommand{\A}{\mathrm{A}}     
\renewcommand{\H}{\mathrm{H}}     
\renewcommand{\P}{\mathrm{P}}     
\newcommand*\chem[1]{\ensuremath{\mathrm{#1}}}
\newcommand{\pos}[1]{\ensuremath{\mathrm{#1}^+}}
\renewcommand{\neg}[1]{\ensuremath{\mathrm{#1}^-}}
\newcommand{\ext}[1]{\ensuremath{\mathrm{#1}^*}}
\begin{document}

\title{Global Simulations of Protoplanetary Disk Outflows
  with Coupled Non-ideal Magnetohydrodynamics and Consistent
  Thermochemistry}

\author{Lile Wang$^{1,2}$, Xue-Ning Bai$^{3}$, and Jeremy
  Goodman$^{1}$}

\footnotetext[1]{Princeton University Observatory,
  Princeton, NJ 08544}
\footnotetext[2]{Center for Computational Astrophysics,
  Flatiron Institute, \\ New York, NY 10010;
  lwang@flatironinstitute.org}
\footnotetext[3]{Institute for Advanced Study, 
  Tsinghua University, \\ Beijing 100084, China;
  xbai@tsinghua.edu.cn
}

\begin{abstract}
  Magnetized winds may be important in dispersing
  protoplanetary disks and influencing planet formation. We
  carry out global full magnetohydrodynamic simulations in
  axisymmetry, coupled with ray-tracing radiative transfer,
  consistent thermochemistry, and non-ideal MHD
  diffusivities.  Magnetized models lacking EUV photons
  ($h\nu>13.6~\eV$) feature warm molecular outflows that
  have typical poloidal speeds $\gtrsim 4~\kms$. When the
  magnetization is sufficient to drive accretion rates
  $\sim 10^{-8}~M_\odot~\yr^{-1}$, the wind mass-loss rate
  is comparable.  Such outflows are driven not centrifugally
  but by the pressure of toroidal magnetic fields produced
  by bending the poloidal field. Both the accretion and
  outflow rates increase with the poloidal field energy
  density, the former almost linearly. The mass-loss rate is
  also strongly affected by ionization due to UV and X-ray
  radiation near the wind base. Adding EUV irradiation to
  the system heats, ionizes, and accelerates the part of the
  outflow nearest the symmetry axis, but reduces the overall
  mass-loss rate by exerting pressure on the wind base.
  Most of our models are non-turbulent, but some with
  reduced dust abundance and therefore higher ionization
  fractions exhibit magnetorotational instabilities near the
  base of the wind.
\end{abstract}

\keywords{accretion, accretion disks ---
  magnetohydrodynamics (MHD) --- planets and satellites:
  formation --- circumstellar matter --- method: numerical }

\section{Introduction}
\label{sec:intro}

Protoplanetary disks (PPDs hereafter) are the birthplaces of
planets. During their $\sim 10^6-10^7~\yr$ lifespans, PPDs
are believed to disperse in three ways: (1) by forming
planets, (2) by accreting onto central protostar, and (3) by
outflowing in winds. The latter two processes compete with
the first, limiting the time and mass available for planet
formation.

Absent magnetic fields, PPD winds must be launched by
photoevaporation. High energy photons heat up the gas by
photoionization/photodissociation, depositing energy into
the gas and unbinding it from the star.

Pioneering theoretical studies of disk photoevaporation
simplified this complex problem in complementary ways.
\citet{2008ApJ...683..287G, 2009ApJ...690.1539G} grafted
analytic Parker winds onto hydrostatic disk models with
relatively detailed thermochemistry.
\citet{2006MNRAS.369..216A, 2006MNRAS.369..229A} performed
hydrodynamic simulations with minimal chemistry and
thermodynamics. \citet{2010MNRAS.401.1415O} took a similar
hydrodynamic approach, using an interpolation table for gas
temperature as a function of ionization parameter.  Recently
\citet{2017ApJ...847...11W} (WG17 hereafter) conducted
hydrodynamic simulations coupled with radiation and
consistent thermochemistry. They found that
photoevaporative mass loss is driven mainly by EUV and
Lyman-Werner FUV photons rather than higher-energy
radiation. \citet{2018ApJ...857...57N} reached similar
conclusions by similar methods and explored the influence of
gas metallicity.

Photoevaporative winds without magnetic fields carry off
only their Keplerian share of angular momentum, and are not
directly relevant to disk accretion. Other processes must be
invoked to explain disk accretion. A magnetized wind,
however, may exert torque on the disk, thereby linking
accretion to outflow. This fact has recently been realized
to be essential as viscous/turbulent accretion via the
magnetorotational instability (MRI,
\citealt{1998RvMP...70....1B}) or other hydrodynamic
instabilities is generally found to be insufficient under
PPD conditions (e.g., \citealt{2013ApJ...769...76B,
  2013ApJ...772...96B, 2013ApJ...764...66S,
  2013ApJ...775...73S}; see \citealt{2014prpl.conf..411T}
for a review).

As is clarified in \citet[hereafter
B17]{2017ApJ...845...75B}, magnetized winds can be divided
into two types: (1) magnetocentrifugal winds, whose poloidal
fields are strong enough to enforce corotation near the wind
base, driving an outflow centrifugally; (2) magneto-thermal
winds driven by the gradient of total pressure, in
particular the energy density in toroidal magnetic fields.
Modeling of global wind kinematics suggests that the second
type prevails in PPDs \citep{2016ApJ...818..152B,
  2016ApJ...821...80B}, but a complete treatment requires
global simulations with consistent microphysics.

The co-evolution of radiation, thermochemistry and non-ideal
MHD in global disk models could be prohibitively
time-consuming if medium or large chemical networks were
applied without proper optimization. MHD simulations to
date have simplified the treatment of microphysics. Most
recently, global simulations have been conducted using
2.5-dimensional axisymmetric full MHD simulations with
non-ideal MHD effects, evaluating magnetic diffusivities by
a pre-calculated interpolation table, and calculating
thermodynamics via a simple relaxation-time recipe with
temperature depending on spatial location
\citep{2017ApJ...845...75B}.

Here we report full axisymmetric MHD simulations coupled
with radiation and thermochemistry: at every time step, in
every spatial zone throughout the simulation domain, a
moderate-scale chemical network (28 species, $\sim 160$
reactions) is evolved.  The network contains species and
reactions that are important for the temperature and
ionization of the gas, and hence for the dynamics of the
flow, but omits some trace species that are important as
observational diagnostics (e.g. neon); these species will be
included in post-processing analyses to be reported in
subsequent publications.  In contrast to simulations using
pre-calculated interpolation tables, our approach offers a
consistent treatment of regions where thermal or chemical
timescales are comparable to flow timescales.

This paper is structured as follows. \S\ref{sec:method}
briefly summarizes the numerical methods and physical
recipes we adopt for carrying out and analyzing our
simulations.  Some details are further elaborated in the
Appendices. \S\ref{sec:fid-model-setup} describes the setup
and parameter choices for our fiducial simulation and
explains the underlying physics, while
\S\ref{sec:fid-results} presents and discusses the main
results for this model. \S\ref{sec:res-param-space} explores
the parameter space to understand the impact of different
physical assumptions. In \S\ref{sec:discussions} we compare
our results to those in B17 and WG17, and discuss
prospective observational tests.  A summary of our main
results and conclusions is given in \S\ref{sec:summary}.

\section{Method}
\label{sec:method}

Our computational methods are summarized in this section. We
describe first the scheme for magnetohydrodynamics (MHD).
Then we discuss our methods for radiative transfer,
thermochemistry, and non-ideal MHD effects, the former being
two largely the same as in WG17.

\subsection{Magnetohydrodynamics}
\label{sec:method-mhd}

We use the grid-based higher-order Godunov MHD code
\verb|Athena++| (\citealt{2016ApJS..225...22W}; J. Stone et
al., in preparation) in spherical-polar coordinates
$(r,\theta,\phi)$. We assume axisymmetry (and hence 2D), and
with reflective reflection symmetry across the equatorial
($\theta = \pi / 2$) plane. All dependence on $\phi$ is
neglected, but $v_\phi$ and $B_\phi$ are still included in
our simulations. For convenience, we also occasionally use
cylindrical coordinates $(R,z)$, defined as
$R\equiv r \sin \theta$, and $z \equiv r \cos \theta$. The
HLLD Riemann solver and piecewise linear reconstruction are
employed in all our simulations.

\verb|Athena++| solves the MHD equations in conservative
form, and therefore conserves mass, energy, (angular)
momentum and magnetic flux to machine precision. The
equations read
\begin{equation}
  \label{eq:mhd}
  \begin{split}
    & \partial_t \rho + \nabla \cdot (\rho \v) = 0\ ;
    \\
    & \partial_t (\rho \v) + \nabla \cdot \left(\rho \v \v -
      \dfrac{\B\B}{4\pi} + P_\tot \mathbf{I} \right) = -
    \nabla \Phi \ ;
    \\
    & \partial_t \B = \nabla \times (\v \times \B - c \E' )\
    ;
    \\
    & \partial_t \epsilon + \nabla \cdot \left[
      \left(\epsilon + P_\tot \right) \v -
      \dfrac{(\B\cdot\v)\B}{4\pi} + \mathbf{S}' \right]
    = 0\ , 
  \end{split}
\end{equation}
where $\rho$, $\v$ and $p$ are the gas density, velocity and
gas thermal pressure respectively, $\B$ is the magnetic
field, $P_\tot\equiv p + B^2/(8\pi)$ is the total pressure,
$\epsilon \equiv p / (\gamma - 1) + \rho (v^2/2 + \Phi) +
B^2/(8\pi)$ is the total energy density ($\gamma$ is the
adiabatic index), $\Phi$ is the gravitational potential, and
$\mathbf{I}$ is the identity tensor.  Non-adiabatic
processes that affect gas energy are calculated separately
in an operator-splitting manner (see
\S\ref{sec:method-micro}).

As a general equation of state (with variable $\gamma$) is
not yet implemented in the current version of
\verb|Athena++| we adopt an ideal equation of state with
constant $\gamma = 5/3$. Admittedly $\gamma$ should increase
from $\gamma=1.4$ in the molecular disk to $\gamma=5/3$ in
regions where molecules are dissociated. We address this
issue further in \S\ref{sec:res-param-space}. 

Non-ideal MHD entails an electric field in the local fluid
rest frame,
\begin{equation}
  \label{eq:non-ideal-def}
  \E' = \dfrac{4\pi}{c^2}
  \left( \eta_\O\J + \eta_\H \J\times \mathbf{b} + \eta_\A
    \J_\bot \right)\ ,
\end{equation}
where $\mathbf{b}\equiv \B/B$ is a unit vector along the
local direction of magnetic field,
$\J = c \nabla \times \B / 4\pi$ is the current density, and
$\J_\bot\equiv \mathbf{b}\times (\J \times \mathbf{b})$ is
the component of $\J$ perpendicular to the local magnetic
field. The Poynting flux associated with $\E'$ reads
$\mathbf{S}' = c\E'\times\B / (4\pi)$. In this paper, we
($\eta_\H=0$).  The Ohmic ($\eta_\O$) and ambipolar
($\eta_\A$) diffusivities are calculated using the
thermochemical network described in
\S\ref{sec:method-micro}.

\subsection{Radiative Transfer and Thermochemistry}
\label{sec:method-micro}

Radiation and thermochemical reactions are important for the
temperature, ionization fraction, and molecular or atomic
composition of the gas.  We solve the coupled
thermochemistry-radiation problem in conjunction with the
MHD equations via operator splitting (see also WG17,
\method).

UV and X-ray photons are emitted isotropically from a point
source at the origin ($r=0$). The luminosity of each ray is
adjusted according to photoreactions and absorptions as it
propagates through the grid cells.  Because individual cells
are sometimes optically thick, we calculate the local
effective flux at photon energy $h\nu$ as
\begin{equation}
  \label{eq:flux-eff}
  F_\eff(h\nu) = \sum_{\{i\mathrm{\ in\ cell}\}} F_i(h\nu)
  \left\{ \dfrac{1 - \exp[-\delta l_i / \lambda(h\nu)]}
    {\delta l_i / \lambda(h\nu)} \right\}\ ,
\end{equation}
where $F_i(h\nu)$ is the flux of the $i$th ray as it enters
the cell, $\delta l_i$ is the chord length of the ray across
the cell, and $\lambda(h\nu)$ is the mean free path of
photon absorption at energy $h\nu$. For each $h\nu$,
$\lambda(h\nu)$ is calculated by collecting all absorption
mechanisms, and is updated as the abundances of species
evolve. As self-/cross-shielding effects are important
especially for some FUV photoreactions, the calculation of
$\lambda(h\nu)$ becomes tricker than for other photon
energies; we refer the reader to Appendix
\ref{sec:appx-shielding} for more details.

In this work we use four discrete energy bins to represent
four important bands of photon energy: $h\nu = 7~\eV$
(``soft FUV'' hereafter, for FUV photons that do not
interact appreciably with hydrogen molecules), $12~\eV$ for
Lyman-Werner (``LW'' ) band photons, $25~\eV$ for EUV
photons, and $3~\keV$ for X-ray photons ( rather than
$1~\keV$ as in WG17).  For FUV and EUV, absorption processes
overwhelm scattering \citep{Verner+Yakovlev1995,
  Verner+etal1996, DraineBook}.  However, scattering may
allow X-ray and $\lya$ photons to penetrate more deeply into
the disk, in part by deflecting radial into latitudinal
propagation.  For X-ray photons, we adopt the recipes
summarized in Appendix \ref{sec:appx-diff-ionize}, together
with recipes for ionization by cosmic rays and radioactive
decays.  \lya photons do not affect the most abundant
chemical species (\chem{H_2}, H, He, CO), yet could still
affect the thermal state by photodissociating \chem{H_2O}
and \chem{OH}, and by photoelectric emission from dust
grains. However, via the same approach as WG17, we have
confirmed that \lya photons have only secondary importance
compared to UV and X-ray photons in both
respects\footnote{Following the scheme in WG17 (based on the
  method described by \citet{2011ApJ...739...78B}), we have
  applied our \lya Monte-Carlo radiation code to our
  fiducial model. We assume that the luminosity in \lya is
  the same as that of the $7~\eV$ soft FUV energy band.  We
  find that (a) in the intermediate layer, the
  photodissociation rate by \lya is at most a tenth of that
  by $7~\eV$ photons; (b) the ionization rate by \lya is
  $\lesssim 10^{-1}$ of that by direct or scattered X-ray
  photons everywhere below the wind base.}. Therefore, in
this paper, we focus on radial ray tracing for the radiative
transfer problem.

Optical and infrared radiation from the central protostar is
not directly relevant to the thermal structure of the
outflow or the ionization structure of the disk (except
within $\sim 0.1\au$ of the star).  Nevertheless, the
temperature profile in the regions unreachable by
high-energy photons is affected by the diffuse radiation in
these two bands, which maintains the dust temperature and
therefore also the gas temperature by thermal accomodation
on grains. Following WG17, we take the following approach.
We set the dust temperature following the simple
  model of \citet{1997ApJ...490..368C}, using the following
equation,
\begin{equation}
  \label{eq:dust-temp}
  \begin{split}
    & 4 \sigb T_\dust^4 \sigma_\dust q(T_\dust) = \\
    & \max \left[ 4 \sigb T_\ah^4(R) \sigma_\dust q(T_\dust)
       \ ,\ \sum_{h \nu} F_\eff(h \nu) \sigma (h \nu)
    \right]\ , 
  \end{split}
\end{equation}
where $T_\ah$ is the desired temperature as a function of
cylindrical radius $R$ (e.g. \citealt{1997ApJ...490..368C},
figure 4),
\begin{equation}
  \label{eq:prof-T_ah}
  T_\ah \simeq 280~\K\times (R/\au)^{-0.5}.
\end{equation}
Also, $\sigma_\dust$ is the geometric dust-grain cross
section, $q(T_\dust)$ is the Planck-averaged emissivity (see
also equation (24.16) in \citealt{DraineBook}),
$F_\eff(h\nu)$ is the local high-energy radiation flux in
photon-energy bin $h \nu$, and $\sigma (h \nu)$ is the
effective absorption cross section. In every timestep,
$T_\dust$ is first calculated by solving
eq. \eqref{eq:dust-temp}. The gas temperature is then
related to the dust temperature via the dust-gas heat
accommodation term \citep{2001ApJ...557..736G,DraineBook},
\begin{equation}
  \label{eq:dust-gas-heat}
  \Gamma_\dust = -  \kb (T -
  T_\dust) \sigma_\dust \sum_\mathrm{sp} n_\mathrm{sp} 
  \left( \dfrac{8 \kb T}{\pi m_\mathrm{sp}} \right)^{1/2}
  \ ,
\end{equation}
where the subscripts ``sp'' range over species. Note that
$\Gamma_\dust$ is negative when the gas is hotter than the
dust.  When other heating/cooling processes are weak, these
prescriptions will stabilize gas temperature at the dust
temperature \eqref{eq:prof-T_ah}.

Advection of chemicals is calculated with special methods
(\method) so that consistency of chemical elements and
species fluxes in every timestep of MHD evolution is
ensured.  Advected abundance of chemicals and internal
energy density is then used as the input to the
thermochemical calculations for the current step.  The
following coupled set of ordinary differential equations
(ODEs) is solved (in which the Einstein summation convention
applies):
\begin{equation}
  \label{eq:ode-chem-thermo}
  \begin{split}
    \dfrac{\d n^i}{\d t} & = \mathcal{A}^i_{\;jk} n^j n^k +
    \mathcal{B}^i_{\;j} n^j\ ; \\
    \dfrac{\d \epsilon}{\d t} & = \Gamma - \Lambda\ ;
  \end{split}
\end{equation}
in which the terms $\{\mathcal{A}^i_{\;jk}\}$ describe
two-body reactions, those in $\{\mathcal{B}^i_{\;j}\}$
represent photoionization, photodissociation, and
spontaneous decays, and $\Gamma$ and $\Lambda$ represent all
non-adiabatic heating and cooling rates per unit volume,
respectively.  Dust-gas heat accomodation
(eq.~\ref{eq:dust-gas-heat}) is included in the $\Gamma$
term.

Because the ODEs \eqref{eq:ode-chem-thermo} are usually
stiff, we use a multi-step implicit method to solve them.
By implementing the thermochemical calculations on GPUs, we
have reduced the (wall-clock) time spent on them so that it
is comparable to that spent by the CPUs on the MHD
(\method).

The set of chemical species and reactions broadly follows
WG17, with some updates so that, based on our tests, the
ionization in the intermediately ionized regions can be
calculated with relatively good accuracy while the cost of
computation is minimized. These 28 species are: \neg{e}
(free electrons), \pos{H}, H, \chem{H_2}, \ext{\chem{H_2}}
[using the $v=6$ vibrational state as a proxy for \chem{H_2}
in all excited states, see \citealt{1985ApJ...291..722T},
(TH85), WG17], He, \pos{He}, O, \pos{O}, OH, \pos{OH},
\chem{H_2O}, C, \pos{C}, CO, \pos{HCO}, \pos{CH}, S,
\pos{S}, \pos{HS}, Si, \pos{Si}, SiO, \pos{SiO}, \pos{SiOH},
Gr, \pos{Gr}, \neg{Gr}. Here Gr and Gr$^\pm$ denote neutral
and singly-charged dust grains, respectively.

The thermochemical mechanisms involved and pertinent
references are summarized below:
\begin{itemize}
\item ``Standard'' two-body interactions in the \verb|UMIST|
  database (\citealt{UMIST2013}; the photochemical reactions
  therein are excluded as they are not suitable for our
  radiation fields).
\item Photoionization/photodissociation of atoms and
  molecules (\citealt{Verner+Yakovlev1995, Verner+etal1996}
  for ionization; TH85 for \chem{H_2} dissociation;
  \citealt{2009A&A...503..323V} for CO dissociation;
  \citet{2014ApJ...786..135A} for OH/\chem{H_2O}
  dissociation); some photoionization/photodissociation
  processes are subject to self-/cross-shielding
  (\citet{2017A&A...602A.105H}; see also Appendix
  \ref{sec:appx-shielding}).
\item Dust-assisted molecule formation
  \citep{Bai+Goodman2009,2014ApJ...786..135A} and
  recombination (\citealt{1987ApJ...320..803D,
    2001ApJS..134..263W}; see also the compilation in
  \citealt{2006A&A...445..205I}); photoelectric emission by
  dust \citep{2001ApJ...554..778L,2001ApJS..134..263W};
  dust-gas heat accommodation (eq. \ref{eq:dust-gas-heat}).
  For simplicity, we follow WG17 and use single-sized dust
  grains with $r_\dust=5~\ang$, which serve as a proxy for
  dust grains of all sizes in terms of their impact on
  thermochemistry and ionization.
\item Cooling by atomic/ionic transitions
  (\citealt{1985ApJ...291..722T}, using an
  escape-probability formalism as described in
  \citealt{1981ApJ...250..478K}); and by ro-vibrational
  transitions of molecules \citep{1993ApJ...418..263N,
    2010ApJ...722.1793O}.
\end{itemize}
Our thermochemical network is tuned to capture the important
reactions relevant to regions near the wind base and in the
wind proper. It is less accurate in the denser regions near
the midplane, particularly with regard to ionization
levels. However, this issue does not undermine our analyses
in this paper: as shown in what follows
(e.g. \S\ref{sec:fid-ang-mom}), the accretion rate is
determined by magnetic stresses near the wind base instead
of those in the midplane.

\subsection{Non-ideal MHD Diffusivities}
\label{sec:method-non-ideal}

Non-ideal MHD diffusivities are determined by the abundances
of charge carriers, which are computed as part of our
thermochemical network. The general expressions for the
three diffusivities read
\citep[e.g.][]{2007Ap&SS.311...35W,2011ApJ...739...51B},
\begin{equation}
  \label{eq:def-diffusivity}
  \begin{split}
    & \eta_\O =
    \dfrac{c^2}{4\pi}\left(\dfrac{1}{\sigma_\O}\right)\ ,\quad
    \eta_\H =
    \dfrac{c^2}{4\pi}
    \left(\dfrac{\sigma_\H}{\sigma_\H^2+\sigma_\P^2}\right)\,
    \quad 
    \\
    & \eta_\A =
    \dfrac{c^2}{4\pi}
    \left(\dfrac{\sigma_\P}{\sigma_\H^2+\sigma_\P^2}\right)
    - \eta_\O\ ,
  \end{split}
\end{equation}
Here $\sigma_\O$, $\sigma_\H$ and $\sigma_\P$ are Ohmic,
Hall, and Pederson conductivities. If $Z_je$ is the charge
and $n_j$ the number density of the $j^{\rm th}$ charged
species, then
\begin{equation}
  \label{eq:def-conductivity}
  \begin{split}
    & \sigma_\O = \dfrac{ec}{B} \sum_j n_jZ_j\beta_j
    \ ,\quad
    \sigma_\H = \dfrac{ec}{B} \sum_j
    \dfrac{n_jZ_j}{1+\beta_j^2}\ ,
    \\
    & \sigma_\P = \dfrac{ec}{B} \sum_j
    \dfrac{n_jZ_j\beta_j}{1+\beta_j^2} ,
  \end{split}
\end{equation}
in which the Hall parameter $\beta_j$ is the ratio of the
gyrofrequency to the collision rate with neutrals,
\begin{equation}
  \label{eq:def-hall-param}
  \beta_j = \dfrac{Z_jeB}{m_jc}
  \dfrac{1}{\gamma_j\rho}\ ;\quad \gamma \equiv
  \dfrac{\mean{\sigma v}_j}{\mean{m}_n + m_j}\ ,
\end{equation}
where $m_j$ is the charged species' molecular mass,
$\mean{m}_n$ is the mean molecular mass of the neutrals, and
$\mean{\sigma v}_j$ is the rate of collisional momentum
transfer between the $j$th species and the neutrals, given
by \citep[e.g.]  [$T_2\equiv (T / 10^2~\K)$] {DraineBook,
  2011ApJ...739...50B},
\begin{equation}
  \label{eq:eval-coll-rate}
  \begin{split}
    & \mean{\sigma v}_e \simeq 8.3\times
    10^{-9}~\cm^3~\s^{-1}\times \max\{T_2^{1/2},1\}\ ,
    \\
    & \mean{\sigma v}_i \simeq 2.0\times
    10^{-9}~\cm^3~\s^{-1}\times \left(
      \dfrac{\mean{m}_nm_i/m_p}{\mean{m}_n+m_i}
    \right)^{-1/2}\ ,
  \end{split}
\end{equation}
for electrons and ions, respectively. As our dust grains are
tiny, their collisional momentum transfer rate
$\mean{\sigma v}_\mathrm{g}$ can be approximated by the rate
for ions, but the actual grain mass is used for $m_j$ in eq.
\eqref{eq:def-hall-param}.  Although all three diffusivities
are calculated, $\eta_\H$ is not actually used in this
paper, as the Hall effect is neglected.

We have tested the diffusivities calculated from our
thermochemical network under the same magnetic, thermal, and
radiation conditions as those of Figure~5 in
\citet{2016ApJ...819...68X}, using a hydrostatic grid and
prescribed temperature.  The error in the Elsasser numbers
$\am$ and $\Lambda_\mathrm{Ohmic}$ is $\lesssim 10\%$ at
higher altitudes ($z/H\gtrsim 2~h_\mid$, where $h_\mid$ is
the gaussian scale height at the equatorial plane), and no
more than one order of magnitude in the mid-plane regions.

The ambipolar diffusivity $\eta_\A$ computed in this way is
very large in two regions that do not much concern us in
this paper.  First, near the polar axis, the density is very
low, the gas is highly ionized ($x_\e\gtrsim 10^{-1}$), and
the one-fluid treatment of magnetic diffusivity simply
fails. Second, near the midplane the ionization is extremely
weak ($x_e\lesssim 10^{-11}$).  To avoid the prohibitively
small time steps that would otherwise be required to evolve
the diffusive terms stably, we cap $\eta_\A$ at
$\eta_{\A,\mathrm{cap}} = [10 c_{s,\mid}
h_\mid]_{r=r_\mathrm{in}}$, where $c_{s,\mid}$ denotes the
adiabatic sound speed at the equatorial plane (similar
schemes capping diffusivity were also adopted in
e.g. \citealt{2013ApJ...769...76B,2015ApJ...801...84G},
B17). This cap does not harm our simulations because (1) in
highly ionized regions, the plasma is already in the ideal
MHD regime; and (2), in regions with very weak ionization,
the magnetic diffusivity is dominated by Ohmic resistivity.
An identical cap is also applied to $\eta_\O$; in
simulations throughout this paper, however, this $\eta_\O$
cap is never reached.

\subsection{Diagnostics of MHD Wind Models}
\label{sec:diag}

This subsection briefly discusses the major diagnostics that
we have used to analyze our simulations.

\subsubsection{Magnetic diffusivities}
\label{sec:diag-mag-diff}

As the Hall effect is neglected, non-ideal MHD effects are
characterized by these two dimensionless Elsasser numbers,
\begin{equation}
  \label{eq:def-elsasser}
  \Lambda_\mathrm{Ohmic} \equiv
  \dfrac{v_\A^2}{\eta_\O\Omega_\K}\ ,\quad 
  \am \equiv \dfrac{v_\A^2}{\eta_\A\Omega_\K}\ ,
\end{equation}
where $v_\A = (B^2/4\pi\rho)^{1/2}$ is the Alfv\'en speed,
and
$\Omega_\K$$=\sqrt{GM_*/r^3}$ is the local Keplerian angular
frequency. Magnetic diffusion is considered to be strong if
either Elsasser number is $\lesssim
1$. In general, $\Lambda_\mathrm{Ohmic} \lesssim
1$ is sufficient to supress MRI \citep[e.g.][]
{2007ApJ...659..729T, 2008A&A...483..815I}.
\citep{2011ApJ...736..144B} found that the plasma $\beta
\equiv 8\pi p / B^2 \lesssim\beta_\min(\am)$ is also able to
damp or suppress MRI, where
\begin{equation}
  \label{eq:mri-beta-min}
  \beta_\min(\am) = \left[
    \left(\dfrac{50}{\am^{1.2}}\right)^2 + 
    \left(\dfrac{8}{\am^{0.3} + 1}\right)^2
  \right]^{1/2}\ .
\end{equation}

\subsubsection{Wind kinematics and dynamics}
\label{sec:diag-wind-kin}

Following B17, we locate the wind base at the FUV front,
where the radial absorption optical depth in the soft FUV
band is unity.  The mechanisms that launch the outflow near
the wind base are of interest. Following
\citet{2016ApJ...818..152B} and B17, we project the forces
acting on fluid elements onto the local poloidal magnetic
field and decompose them into three terms,
\begin{equation}
  \label{eq:force-decomp}
  \begin{split}
    & \left( \dfrac{\d v_\p}{\d t} \right)_s
    = f_\mathrm{gas} + f_\mathrm{ine} + f_\mathrm{mag}\ ;
    \\
    & f_\mathrm{gas} \equiv -\dfrac{\d p}{\d s}\ ;
    \quad
    f_\mathrm{ine} \equiv \dfrac{v_\phi^2}{R}
    \dfrac{\d R}{\d s} - \dfrac{\d \Phi}{\d s}\ ;
    \\
    & f_\mathrm{mag} = -\dfrac{B_\phi}{4\pi\rho R}
    \dfrac{\d (R B_\phi)}{\d s}\ ,
  \end{split}
\end{equation}
where, along a field line, $f_\mathrm{gas}$ and
$f_\mathrm{mag}$ are accelerations due to gas and magnetic
pressure gradients respectively, and $f_\mathrm{ine}$ is the
net inertial acceleration (combining centrifugal and
gravitational forces).

The wind mass loss rate $\dot{M}_\wind$ is an important
diagnostic of disk dispersal. We characterize
$\dot{M}_\wind$ by the local wind mass loss rate per
logarithmic radius,
\begin{equation}
  \label{eq:def-modt-lnr}
  \dfrac{\d\dot{M}_\wind}{\d\ln R} = 4\pi R^2 \mean{\rho
    v_z}|_{z_\wb}\ .
\end{equation}
We write $\dot M_\wind$ for the integral of this between our
radial boundaries.

Following B17, we use the following quantities that would be
conserved along magnetic field lines in steady winds in the
ideal MHD regime,
\begin{equation}
  \label{eq:hydro-mag-diag}
  \begin{split}
    & k \equiv \dfrac{4\pi \rho v_\p}{B_\p}\, ,\quad \omega
    \equiv \dfrac{v_\phi}{R} - \dfrac{k B_\phi}{4\pi \rho
      R}\, ,\quad
    l \equiv v_\phi R - \dfrac{R B_\phi}{k}\, ;
    \\
    & k_0 \equiv \dfrac{4\pi
        \rho_\mid v_\K}{B_{z0}}\ ,\quad
    \omega_0 \equiv \Omega_{\K}(R_\wb)
    \, ,
    \\
    & l_0 \equiv \Omega_{\K}(R_\wb)R_\wb^2\ .
  \end{split}
\end{equation}
Here the subscript ``p'' denotes a poloidal component,
$v_\K$ is the Keplerian speed, and $R_\wb$ is the
cylindrical radius of the wind base along the same field
line. The symbols $k$, $\omega$ and $l$ stand for poloidal
mass flux per magnetic flux, angular velocity of the
magnetic line, and specific angular momentum, while $k_0$,
$\omega_0$ and $l_0$ are reference values defined at or near
the wind base.

\subsubsection{Wind versus accretion: Angular momentum
  transfer}
\label{sec:diag-wind-acc}

A key concept of all steady magnetized wind models is the
poloidal Alfv\'en radius $R_\A$, which is defined as the
cylindrical radius of the point along the field line where
the poloidal velocity $v_\p$ equals the poloidal Alfv\'en
velocity $v_{\A,\p}\equiv B_\p/(4\pi\rho)^{1/2}$. At $R_\A$,
the accretion and outflow rates are related 
in steady state by
\citep[e.g.][]{1995A&A...295..807F,2016ApJ...818..152B}
\begin{equation}
  \label{eq:wind-acc-rate}
  \xi \equiv
  \left(\dfrac{1}{\dot{M}_\acc} \right)  
  \left(\dfrac{\d \dot{M}_\wind}{\d\ln R} \right)
  = \dfrac{1}{2} \dfrac{1}{(R_\A/R_\wb)^2 - 1}\ .
\end{equation}
The quantity $(R_\A/R_\wb)$ is often referred to as the
``magnetic lever arm,'' and $\xi$ is called the ``ejection
index.''

In a steady or statistically states where magnetic stresses
dominate the transport of angular momentum,
\begin{equation}
  \label{eq:acc-stress}
  \dfrac{\dot{M}_\acc v_\K}{4\pi} =
  \dfrac{\partial}{\partial R} \left( R^2
    \int^{z_\wb}_{-z_\wb}\d z\ \mean{T_{R\phi}}_{t,\phi}
  \right)
  + \left. R^2 \mean{T_{z\phi}}_{t,\phi}
  \right|^{z_\wb}_{-z_\wb}\ .
\end{equation}
Here
$\dot{M}_\acc \equiv -2\pi R \int^{z_\wb}_{-z_\wb}\d z\ \rho
v_R$ is the radial accretion rate at radius $R$,
$T_{ij}\equiv -B_iB_j/(4\pi)$ are components of the Maxwell
stress tensor (not to be confused with gas temperature), and
angle brackets stand for temporal and azimuthal
averaging. Eq. \eqref{eq:acc-stress} assumes that azimuthal
velocities are close to the local Keplerian speed. The first
term in eq. \eqref{eq:acc-stress} resembles a radial viscous
stress and can be characterized by the classic dimensionless
$\alpha$ parameter \citep{1973A&A....24..337S},
\begin{equation}
  \label{eq:def-alpha}
  \alpha \equiv \left[ \int_{-z_\wb}^{z_\wb}\d z\ p
  \right]^{-1}\times
  \int_{-z_\wb}^{z_\wb}\d z\ \mean{T_{R\phi}}\ . 
\end{equation}
The second term in eq. \eqref{eq:acc-stress} represents the
vertical component of angular momentum transfer. Comparison
of these two terms reveals which type of angular momentum
transport mechanism dominates in the disk. 

\subsubsection{Energy budget}
\label{sec:diag-ene-budget}

Analyzing the energy budget of the wind helps us to compare
various factors that contribute to wind dynamics and
thermodynamics. In steady states ($\partial_t\rightarrow 0$)
the rates of change of the gas internal and mechanical energy
read, respectively \citep[see also][] {1998RvMP...70....1B},
\begin{equation}
  \label{eq:ene-budget-local}
  \begin{split}
    & \nabla \cdot F_{\mathrm{int}} = - p\nabla\cdot \v +
    \Gamma - \Lambda + (\Gamma_\mathrm{A} +
    \Gamma_\mathrm{O}) \ ;
    \\
    & \nabla\cdot F_{\mathrm{mech}} = p \nabla\cdot\v +
    \nabla\cdot F_{\mathrm{mag}}  - (\Gamma_\mathrm{A}
    + \Gamma_\mathrm{O})\ ,
  \end{split}
\end{equation}
where $\Gamma$ and $\Lambda$ are thermal chemical heating
and cooling rates of the gas
(eq.~\ref{eq:ode-chem-thermo}), and
\begin{equation}
  \Gamma_\mathrm{A} \equiv \dfrac{4\pi\eta_\mathrm{A}}{c^2}
  \J_\bot^2\ ,\quad
  \Gamma_\mathrm{O} \equiv \dfrac{4\pi\eta_\mathrm{O}}{c^2}
  \J^2\ ,
\end{equation}
are the heating rates due to ambipolar diffusion and Ohmic
resistivity, and
\begin{equation}
   F_{\mathrm{int}} \equiv \dfrac{p\v}{\gamma - 1}\ ,\quad
    F_{\mathrm{mech}} \equiv \left(\dfrac{\rho v^2}{2} + \rho
      \Phi + p\right) \v\ ,
\end{equation}
are the fluxes in internal and mechanical energy of the gas,
and
\begin{equation}
  F_{\mathrm{mag}} \equiv - \dfrac{\B\times
      (\v\times\B)}{4\pi} - \mathbf{S}'\ ,
\end{equation}
denotes the dynamic flux of electromagnetic
fields. Decomposing $\nabla\cdot F_\mathrm{int}$ and
$\nabla \cdot F_\mathrm{mech}$ by
eq.~\eqref{eq:ene-budget-local} will shed light onto the
energetic evolution of fluids at different spacial
locations.

In addition to the local energy balance
(eq.~\ref{eq:ene-budget-local}), another flavor of energy
budget analysis involves the wind as a whole. Integrating
the combination of eqs.~\eqref{eq:ene-budget-local} on the
volume of a steady wind, we obtain,
\begin{equation}
  \label{eq:ene-buget-global}
  \dot{E}_{\mathrm{int}} +  \dot{E}_{\mathrm{mech}} 
  + \dot{E}_{\mathrm{na}} + \dot{E}_{\mathrm{mag}} = 0\ .
\end{equation}
Our sign convention is that $\dot{E}_X > 0$ means that
energy of type $X$ leaves the wind region (for example,
$\dot{E}_\mathrm{cool} \geq 0$ in the following
eq~\ref{eq:ene-local-cool-heat}).  Using the divergence
theorem, the internal term quantifies the net power brought
into the wind by internal energy flux then reads,
\begin{equation}
  \dot{E}_{\mathrm{int}} \equiv
  \oint\limits_{\partial(\wind)} \d \hat{S} \cdot 
  F_\mathrm{int}\ .
\end{equation}
Here the subscript ``$\partial(\wind)$'' denotes the
boundaries of the wind region, and $\d\hat{S}$ is the area
integration measure with outward-pointing normal. The
mechanical term is decomposed into the net power brought by
fluxes in kinetic (further decomposed into toroidal and
poloidal components), gravitational and $p\d V$ energy,
\begin{equation}
  \begin{split}
    & \dot{E}_{\mathrm{mech}} \equiv
    \dot{E}_{\mathrm{k,p}} + \dot{E}_{\mathrm{k,}\phi} +
    \dot{E}_{\mathrm{grav}} + \dot{E}_{p\d V} \ ;
    \\
    & \dot{E}_{\mathrm{k,p;}\phi} \equiv
    \oint\limits_{\partial(\wind)} \d \hat{S} \cdot \v
    \left(\dfrac{\rho v_{\p;\phi}^2}{2} \right)\ ,\
    \\
    & \dot{E}_{\mathrm{grav}} \equiv
    \oint\limits_{\partial(\wind)} \d \hat{S} \cdot \v
    \rho \Phi\ ,\ \dot{E}_{p\d V} \equiv
    \oint\limits_{\partial(\wind)} \d \hat{S} \cdot \v p\ .
  \end{split}
\end{equation}
The non-adiabatic term integrates contributions of
thermochemical processes over the wind volume,
\begin{equation}
  \label{eq:ene-local-cool-heat}
  \dot{E}_{\mathrm{na}} \equiv \dot{E}_{\mathrm{heat}}
  + \dot{E}_{\mathrm{cool}} \equiv
  \int\limits_\wind \d V\ (-\Gamma)
  + \int\limits_\wind \d V\ \Lambda\ .
\end{equation}
The power in electromagnetic fields sums up the
contributions of magnetic stress, magnetic energy flux and
Poynting flux,
\begin{equation}
  \label{eq:ene-budget-em}
  \begin{split}
    & \dot{E}_{\mathrm{mag}} \equiv \dot{E}_{\mathbf{S}'} +
    \dot{E}_{\B} + \dot{E}_{\mathrm{stress}} \ ;
    \\
    & \dot{E}_{\mathbf{S}'}
    \equiv \oint\limits_{\partial(\wind)} \d \hat{S} \cdot
    \mathbf{S}'\ ,\ 
    \dot{E}_{\mathbf{B}} \equiv
    \oint\limits_{\partial(\wind)} \d \hat{S} \cdot
    \dfrac{\v B^2}{4\pi}\ ,
    \\
    & \dot{E}_{\mathrm{stress}} =
    \oint\limits_{\partial(\wind)} \d \hat{S} \cdot
    \left[-\dfrac{(\B\cdot\v)\B}{4\pi} \right] \ .
  \end{split}
\end{equation}
We notice that the power of magnetic stress term is usually
dominated by the power at which magnetic stress does work on
the wind base,
\begin{equation}
  \dot{E}_{\mathrm{stress}} \simeq 
  -\int\limits_\wb \d S\ \dfrac{B_zB_\phi v_\phi}{4\pi}\ .
\end{equation}

\section{Fiducial Model Setup}
\label{sec:fid-model-setup}

\subsection{ Geometry of the Simulation Domain}
\label{sec:fid-geometry}

The simulations are axisymmetric (independent of $\phi$),
spanning the radial range $(r/\au)\in [1,100]$ and
colatitude range $\theta \in [0.035,\pi/2]$.  Reflection
symmetry about the equatorial plane (midplane:
$\theta=\pi/2$) is imposed; all global quantities (e.g. wind
mass loss rate, accretion rate) include the contributions
from both sides of the plane. The standard resolution is 384
radial by 128 latitudinal. The radial zones are spaced
logarithmically. The latitudinal zones have grid spacing
$\Delta \theta$ decreasing in geometric progression from
pole to mid-plane, so that $\Delta \theta$ at the midplane
is $1/6$ as large as near the pole. As the mid-plane scale
height is roughly $h_\mid\sim 0.05-0.10 R$, this grid
geometry gives 20-25 latitudinal zones per $h_\mid$ at the
disk mid-plane.

\subsection{Central Protostar and Radiation Sources}
\label{sec:fid-centroid}

The gravitational field is provided by a 1~$M_\odot$ point
source located at the origin. The radiation sources are
co-centered at the origin and radiate rays isotropically
into the simulation domain. In order to approximate
attenuation of radiation from the source interior to the
inner boundary, we adopt a ``pre-absorption'' recipe that is
elaborated in Appendix \ref{sec:appx-pre-absorb}. This
avoids instabilities caused by unphysically high ionization
rates due to unattenuated rays reaching the mid-plane near
the inner boundary.  In the fiducial model, the EUV bin is
{\it turned off}, for better comparison with B17. The number
of photons radiated in soft FUV and LW FUV bands per unit
time follows a $9000~\K$ black body integrated over the
photon energy range $6~\eV<h\nu < 13.6~\eV$, with total
luminosity $L_\mathrm{FUV} = 10^{31.7}~\erg~\s^{-1}$ (see
also \citealt{2009ApJ...690.1539G}, WG17). The X-ray
luminosity is $L_X=10^{30}~\erg~\s^{-1}$, matching B17 but
approximately half of the value in WG17.

\subsection{Initial and Boundary Conditions}
\label{sec:fid-init-bound}

The initial disk density and temperature profiles follow the
steady state solution in \citet{2013MNRAS.435.2610N}: we set
the mid-plane density as
$\rho = 2.38\times 10^{14}~m_p~\cm^{-3}~(R/\au)^{-q_\rho}$
and temperature $T=280~\K~(R/\au)^{-q_T}$, with the 
radial power indices $q_\rho = 2.25$ for density and
$q_T=0.5$ for temperature (note that, at the mid-plane, this
initial temperature profile is identical to the prescribed
dust temperature profile in eq. \ref{eq:prof-T_ah}). The
disk mass is $\simeq 0.033~M_\odot$ within $50\au$.  Such
hydrodynamic initial conditions generally follow B17;
compared to WG17, this disk has $\sim 1.5\times$ the
midplane scale height.

The initial abundances of the chemical species are set
uniformly throughout the domain according to Table
\ref{table:fiducial-model} ($n_{\chem{H}}$ is the number
density of hydrogen nuclei).  The abundances of elements
generally follow a subset of those in
\citet{2008ApJ...683..287G, 2009ApJ...690.1539G} (note that
the elemental abundances of S and Si are subject to
depletion compared to the Solar abundances; see also
\citealt{2009ApJ...700.1299J}), with the additional
assumption that elements appear in chemical compounds if
possible.

The initial poloidal fields are described by a purely
azimuthal vector potential
$\mathbf{A} \equiv A_\phi \hat{\phi}$,
\citep{2007A&A...469..811Z},
\begin{equation}
  \label{eq:init-vec-potential}
  A_\phi = \dfrac{B_0r_0}{4 - \alpha - q_T}
  \left(\dfrac{R}{r_0}\right)^{1-\frac{\alpha+q_T}{2}}
  \left[ 1 + (m \tan\theta)^{-2} \right]^{-5/2}\ ,
\end{equation}
where $r_0$ is some reference radius (we use $r_0=1~\au$),
and $B_0$ is the midplane field intensity at $r_0$, which is
controlled by the plasma $\beta_0$$\equiv 8\pi
p_0/B_0^2$, $p_0$ being the gas pressure at the midplane.
Following B17, we choose $\beta_0=10^5$ for the fiducial
model, which approximately yields disk accretion rate $\sim
10^{-8}~M_\odot~\yr^{-1}$, 
\begin{equation}
  \label{eq:init-b-midplane}
  B_\p(\theta = \pi / 2 ) = B_0
  \left(\dfrac{R}{r_0}\right)^{-\frac{\alpha+q_T}{2}}\ .
\end{equation}
In the vector potential \eqref{eq:init-vec-potential},
$m$ is a control parameter that describes the bending of
these initial poloidal fields; here we choose
$m=1$.  As did B17, we have verified that reasonable
variations of
$m$ do not affect the final quasi-steady state.

The boundary conditions deserve special attention in order
to avoid unphysical results. As the innermost domain of
$r<1~\au$ is not included by our simulations, any fluxes
(fluid, magnetic) emerging from the inner radial boundary
should not affect the simulation domain.  Since the standard
outflow boundary conditions could violate causality and make
the system unstable, we adopt boundary conditions similar to
B17.  Within the ghost zones for the inner boundary, the
density and velocity are set according to the steady state
solution used for the initial conditions, while the gas
temperature and relative abundances of chemical species copy
the values in the innermost radial zone. The magnetic fields
are extrapolated from the innermost radial zone assuming
$B_r\propto r^{-2}$, $B_\phi\propto
r^{-1}$, and $B_\theta\propto
r^0$.  Inside the simulation domain we also set a radial
buffer zone, $r\in [r_\mathrm{in},
1.5r_\mathrm{in}]$, in which the MHD diffusivities are
linearly tapered to zero as the radius decreases to
$r_\mathrm{in}$, and the poloidal velocity is damped at the
local orbital timescale. This buffer zone helps to stabilize
our simulations near the inner radial boundary. On the outer
radial boundary, outflow boundary conditions (with inflow
inhibitor) are applied to the fluid variables, while the
scheme for magnetic fields is the same as at the inner
radial boundary. The latitudinal boundaries are reflective:
they keep the normal component of
$\B$ and flip tangential components.

\subsection{Simulation Run}
\label{sec:fid-run}

To relax hydrodynamic and thermochemical transients, we run
the simulation for $2000~\yr$, from $t=-2000~\yr$ to $t=0$,
with hydrodynamics, radiation, and thermochemistry, but
without MHD.  Instantaneously, at $t=0$, non-ideal MHD is
turned on, and an external poloidal magnetic field described
by eq. \eqref{eq:init-vec-potential} is applied to the
system. After that, this model is evolved to $t=2000~\yr$,
by which time it has usually reached a quasi-steady state,
except in unstable cases such as Model~7 (see below).

\begin{deluxetable}{lr}
  \tablecolumns{2} 
  \tabletypesize{\scriptsize}
  \tablewidth{0pt}
  \tablecaption{Properties of the fiducial model
    \label{table:fiducial-model}
  }
  \tablehead{
    \colhead{Item} &
    \colhead{Value}
  }
  \startdata
  Radial domain & $1~\au \le r \le \ 100~\au$\\
  Latitudinal domain & $0.035\le\theta\le\pi/2$ \\
  Resolution & $N_{\log r} = 384$, $N_\theta= 128$ \\
  \\
  Stellar mass & $1.0~M_\odot$ \\
  \\
  $M_\disk(1~\au \leq r \leq 100~\au)$ & $0.033~M_\odot$
  \\[2pt] 
  Initial mid-plane density &
  $2.38\times 10^{14}(R/\au)^{-2.25}~m_p~\cm^{-3}$ \\[2pt]
  Initial mid-plane plasma $\beta$ & $10^5$ \\
  Initial mid-plane temperature &
  $280(R/\au)^{-0.5}~\K$ \\
  Artificial heating profile &
  $280(R/\au)^{-0.5}~\K$ \\
  \\
  Luminosities [photon~$\s^{-1}$] & \\[5pt]
  $7~\eV$ (``soft'' FUV)  & $4.5\times 10^{42}$ \\
  $12~\eV$ (LW) & $1.6\times 10^{40}$ \\ 
  $25~\eV$ (EUV) & $0$ \\
  $3~\keV$ (X-ray) & $2.1\times 10^{38}$ \\
  \\
  Initial abundances [$n_{\chem{X}}/n_{\chem{H}}$] & \\[5pt]
  \chem{H_2} & 0.5\\
  He & 0.1\\
  \chem{H_2O} & $1.8 \times 10^{-4}$\\
  CO & $1.4 \times 10^{-4}$\\
  S  & $2.8 \times 10^{-5}$\\
  SiO & $1.7 \times 10^{-6}$\\
  Gr & $1.0 \times 10^{-7}$ \\
  \\
  Dust/PAH properties & \\
  $r_\dust$ & $5~\ang$ \\
  $\rho_\dust$ & $2.25~\g~\cm^{-3}$ \\
  $m_\dust/m_\mathrm{gas}$ & $7\times 10^{-5}$ \\
  $\sigma_\dust/\chem{H}$ & $8\times 10^{22}~\cm^2$ \\
  \enddata
\end{deluxetable}

\section{Fiducial Model Results}
\label{sec:fid-results}

\begin{figure*}
  \centering
  \includegraphics[width=7.0in, keepaspectratio]
  {\figdir/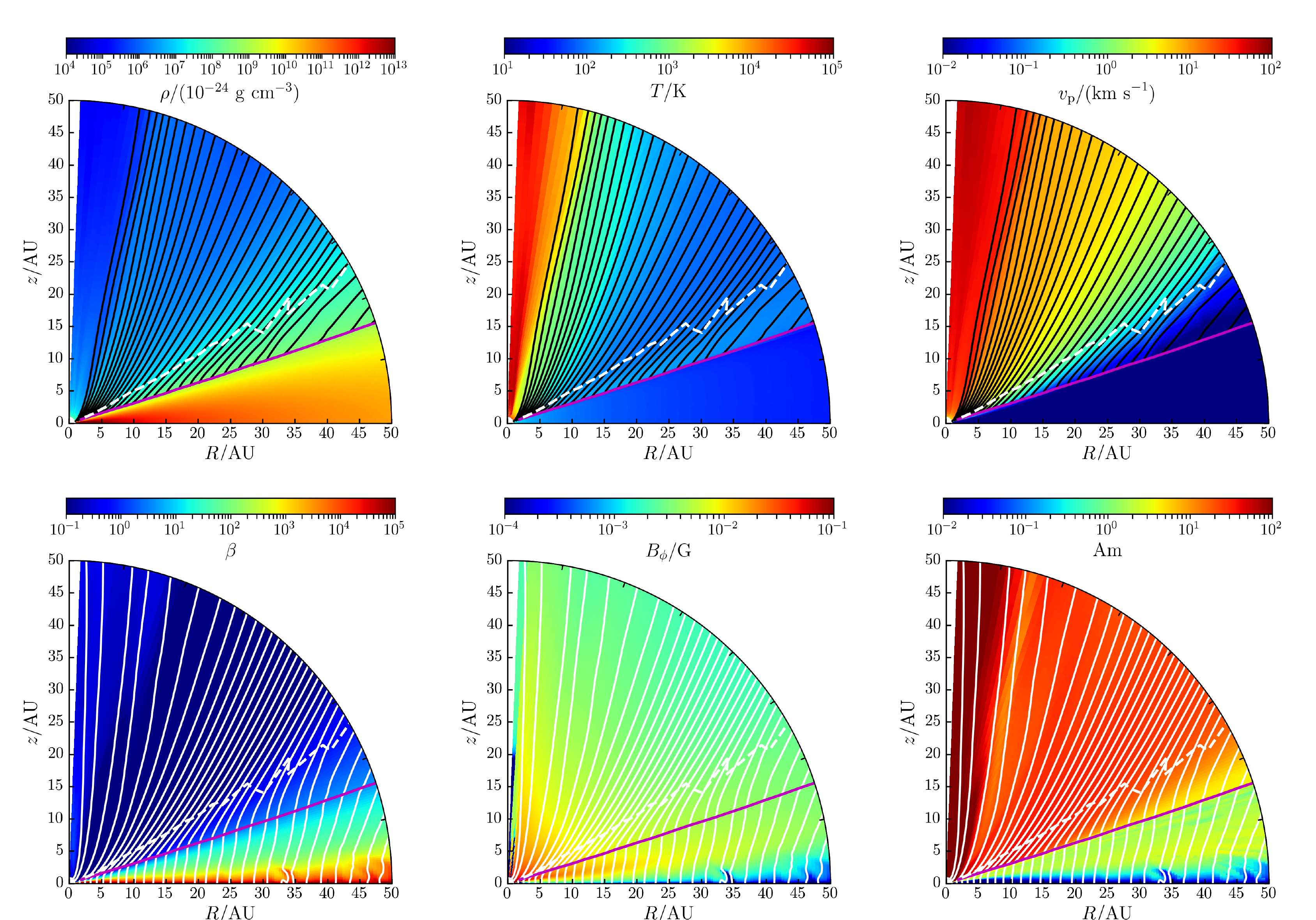}
  \caption{Meridional plots for the fiducial model (Model 0;
    \S\ref{sec:fid-model-setup}), showing the innermost
    $50~\au$ averaged over the final $50~\yr$.  {\bf Top
      row:} basic hydrodynamic profiles. {\it Left panel:}
    mass density in units of $10^{-24}~\g~\cm^{-3}$; {\it
      middle panel:} temperature in Kelvin; {\it right
      panel:} poloidal velocity in units of $\kms$.  {\bf
      Bottom row:} MHD-related profiles.  {\it Left panel:}
    plasma $\beta$ ; {\it middle panel:} toroidal magnetic
    field in units of Gauss; {\it right panel:} ambipolar
    Elsasser number $\am$ [eq.~\eqref{eq:def-elsasser}].
    {\it Black curves, upper panels:} poloidal streamlines
    separated by $10^{-9}~M_\odot~\yr^{-1}$.  {\it White
      solid curves, lower panels:} poloidal magnetic field
    lines.  {\it White dashed curves:} poloidal Alfv\'enic
    surface.  {\it Magenta lines:} wind bases (defined as
    the FUV front, \S\ref{sec:diag-wind-kin}).}
  \label{fig:fiducial_slice}
\end{figure*}

Figure~\ref{fig:fiducial_slice} exhibits meridional plots of
the fiducial model (Model 0) averaged over the final
$50~\yr$ (from $t=1950~\yr$ to $2000~\yr$) of the
simulation, during which the model has already reached
quasi-steady state: $\mean{\dot{M}_\wind}_{50}$ (the
$50~\yr$ average of wind mass loss rate) exhibits less than
$\sim 5~\%$ r.m.s. variation during $10^3~\yr$.
Black solid curves in the top and bottom row are
streamlines, integral curves of the poloidal vector field
$\rho v_\p$ spaced by constant wind mass loss rate
$10^{-9}~M_\odot~\yr^{-1}$. White solid curves in the bottom
row are magnetic field lines spaced by constant poloidal
magnetic flux $5\times 10^{25}~\G~\cm^2$. Mass fluxes
between neighboring stream lines are integrated over azimuth
and are multiplied by two, to include both sides of the
equatorial plane. The fluid streamlines are masked out below
the wind base.

A stream or field line that originates from the disk at
cylindrical radius $R_0 > 50~\au$ may not reach its poloidal
Alfv\'enic point before it leaves the simulation domain at
$r=100~\au$. Furthermore, the outer regions take longer to
reach steady state, often longer than $2000~\yr$.
Therefore, in what follows, our analyses will be limited to
$r<50~\au$.

\begin{figure*}
  \centering
  \includegraphics[width=7.0in, keepaspectratio]
  {\figdir/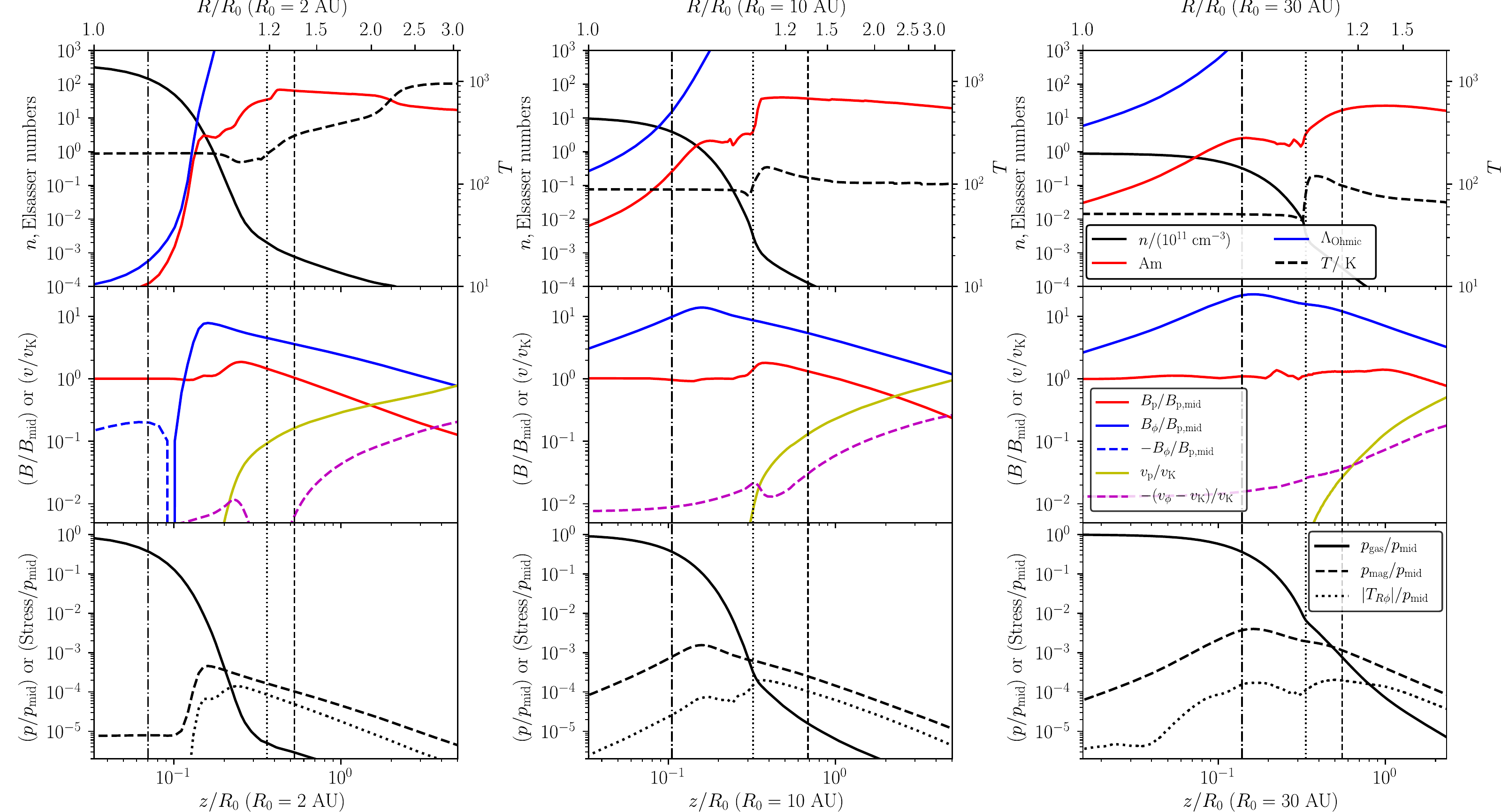}
  \caption{ Flow profiles of the fiducial model (Model 0),
    averaged over the final $50~\yr$, along three different
    magnetic field lines, which intercept the equatorial
    plane at cylindrical radii $R_0=2~\au$ {\it (left
      column),} $R_0=10~\au$ {\it (middle column),} and
    $R_0=30~\au$ {\it (right column)}.  Normalized
    coordinates $(R/R_0,\,z/R_0)$ along field lines shown on
    upper \& lower abscissae.  Profiles are distinguished by
    line shape and color, as marked by the legends.
    Vertical dash-dotted, dotted and dashed lines in black
    indicate the midplane scale height ($h_\mid$), the wind
    base ($z_\wb$, the soft FUV front;
    \S\ref{sec:diag-wind-kin}), and the poloidal Alfv\'enic
    point, respectively.
    \\
    {\it First row:} number density of hydrogen nuclei $n$,
    Ohmic \& ambipolar Elsasser numbers
    $\Lambda_\mathrm{Ohmic}$ \& $\am$
    [eq.~\eqref{eq:def-elsasser}], and temperature $T$
    (right ordinate).  {\it Second row:} poloidal ($B_p$)
    and toroidal ($B_\phi$) components of the magnetic field
    (normalized to midplane value on the current field line,
    $B_{\rm p,\mid}$), and poloidal ($v_\p$) and toroidal
    ($v_\phi$) components of fluid velocity (normalized to
    the Keplerian velocity at the {\it local} cylindrical
    radius, $v_\K\equiv \sqrt{GM_*/R}$).  {\it Third row:}
    gas pressure ($p_\mathrm{gas}$), magnetic pressure
    ($p_\mathrm{mag}$) and $(R,\phi)$ component of Maxwell
    stress tensor ($T_{R\phi}\equiv -B_RB_\phi/4\pi$),
    normalized by $p_\mid\equiv p_\mathrm{gas}|_{z=0}$.  }
  \label{fig:fiducial_vert}
\end{figure*}

In quasi-steady states, the outflow typically becomes
super-Alfv\'enic within a factor or two or less in
cylindrical radius beyond the wind base. Hence the outflow
is a wind, rather than a sub-Alfv\'enic breeze; in what
follows we will use the terms ``wind''/``outflow''
interchangeably unless specifically noted.  If we integrate
the radial outflow above wind bases over the sphere at
$r=50~\au$, excluding streamlines emanating from the inner
radial boundary, we obtain a total wind mass loss rate of
$\dot{M}_\wind \simeq 2.9 \times 10^{-8}~M_\odot~\yr^{-1}$.
The mean radial outflow velocity weighted by radial mass
flux is $\mean{v_r}\simeq 4.6~\kms$. In the mid-plane, the
radial accretion rate is almost constant with radius,
varying slightly within the range
$1.3-2.1\times 10^{-8}~M_\odot~\yr^{-1}$.  For the same
$50~\yr$ average, Figure~\ref{fig:fiducial_vert} displays
the run of several flow variables along three representative
magnetic field lines with footpoints on the equatorial plane
at $R_0=2~\au$, $10~\au$ and $30~\au$.  These plots will
inform our analysis of the physics of the wind in the rest
of this section.

\subsection{Magnetohydrodynamics of Disk and Wind}
\label{sec:fid-dyn}

\subsubsection{Wind launching and kinematics}
\label{sec:fid-wind-launch}

How is the disk wind launched?  Before the poloidal magnetic
field is applied, the fiducial simulation has been run for
$2000~\yr$ (\S\ref{sec:fid-run}), during which no
appreciable outflow was observed
($\dot{M}_\wind < 10^{-11}~M_\odot~\yr^{-1}$)---similar to
the model without EUV in WG17.  Evidently, magnetic fields
are crucial for launching the wind, at least with our
fiducial parameters (which omit EUV).

\begin{figure}
  \centering
  \includegraphics[width=3.0in, keepaspectratio]
  {\figdir/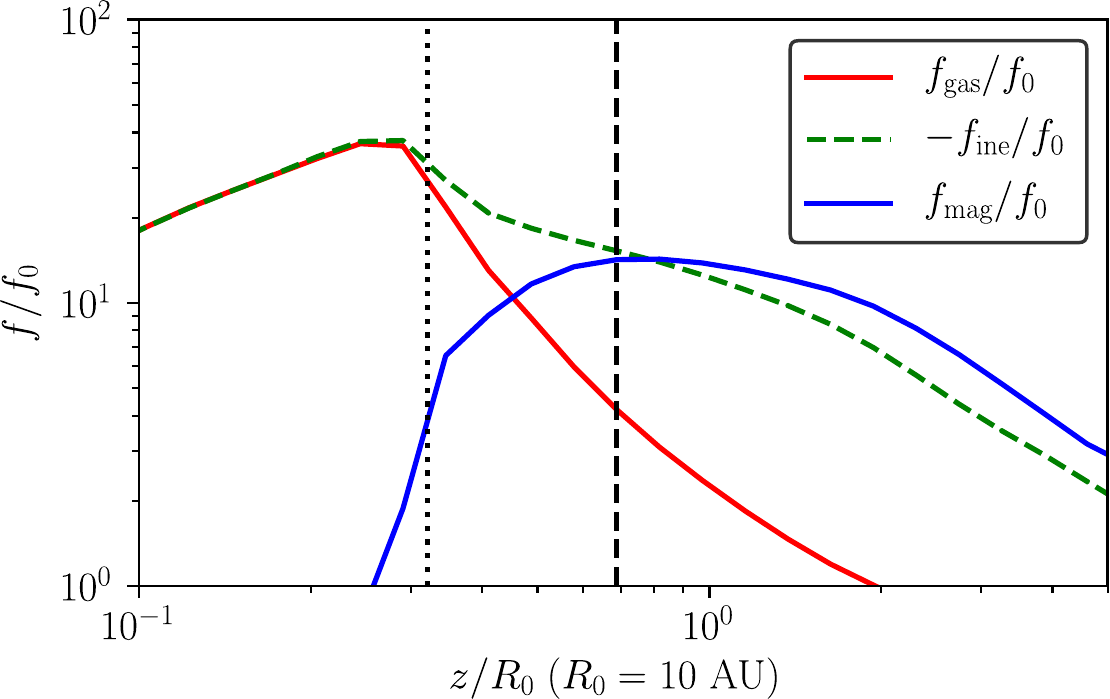}
  \caption{
    Poloidal forces
      [eq.~\eqref{eq:force-decomp}] along the field line
      anchored at $R_0=10~\au$, averaged over the final
      50~yr of Model~0. All forces are normalized to
    $f_0\equiv p_\mid / (R_0 \rho_\mid)$.  Vertical dotted
    and dashed lines 
 mark the wind base and 
    poloidal Alfv\'enic point, respectively. }
  \label{fig:fiducial_force}
\end{figure}

Figure~\ref{fig:fiducial_force} illustrates the force
decomposition defined by eq.~\eqref{eq:force-decomp} along
the field line originating from $R_0=10~\au$, which we find
representative for field lines at all radii. We note that
the net inertial acceleration $f_\mathrm{ine}$ is negative
everywhere. In fact, the toroidal velocity $v_\phi$ is
everywhere sub-Keplerian (see the second row of
Figure~\ref{fig:fiducial_vert}). The magnetic field is too
weak ($\beta_{\rm wind}\gtrsim 10^{-1}$) to enforce
co-rotation, as would be required for magnetocentrifugal
acceleration \citep[see also ][]
{2016ApJ...818..152B}. Instead, the poloidal lines wind up
into toroidal ones under the inertia of the fluid, and the
combined thermal and magnetic pressures launch the wind.
B17 has dubbed such winds ``magneto-thermal.''

Above the wind base, toroidal fluid velocities start to
deviate appreciably from the local $v_\K$, and poloidal
components become important, increasing to
$v_\p\sim 10^{-1}v_\K$.  Figure~\ref{fig:fiducial_cons_diag}
displays the diagnostic quantities \eqref{eq:hydro-mag-diag}
along the $R_0=10\au$ field line. These quantities are
roughly but not strictly conserved, (e.g., field lines
deviate from fluid streamlines, as is seen in
Figure~\ref{fig:fiducial_slice}). We notice that the
specific angular momentum $l$ is approximately
$1.7~\Omega_\K(R_0)R_0^2$: a fluid element on this field
line carries off only 70\% more than its original Keplerian
angular momentum as it flies to infinity, implying that the
local wind mass loss rate should be comparable to the
accretion rate.

\subsubsection{Wind energy budget}
\label{sec:fid-wind-ene-budget}

\begin{deluxetable}{lr}
  \tablecolumns{2} 
  \tabletypesize{\scriptsize}
  \tablewidth{200pt}
  \tablecaption{Model 0 wind energy budget.
    \label{table:fid-ene-budget}
  } 
  \tablehead{
    {Item} &
    \colhead{Value [$10^{30}~\erg~\s^{-1}$]}
  }
  \startdata
  $\dot{E}_{\mathrm{mech}}$ & $1.40$ \\ [2pt]
  \qquad $\dot{E}_{\mathrm{k,p}}$ & $0.40$ \\
  \qquad $\dot{E}_{\mathrm{k},\phi}$ & $-1.06$ \\
  \qquad $\dot{E}_{\mathrm{grav}}$ & $2.05$ \\
  \qquad $\dot{E}_{p\d V}$ & $0.01$ \\ [5pt]
  $\dot{E}_{\mathrm{mag}}$ & $-1.53$ \\ [2pt]
  \qquad $\dot{E}_{\mathrm{stress}}$ & $-1.60$ \\
  \qquad $\dot{E}_{\mathbf{S}'}$ & $0.02$ \\
  \qquad $\dot{E}_{\B}$ &  $0.05$ \\[5pt]
  $\dot{E}_{\mathrm{int}}$ & $0.02$ \\[5pt]
  $\dot{E}_{\mathrm{na}}$ & $0.10$ \\ [2pt]
  \qquad $\dot{E}_{\mathrm{heat}}$ & $-0.10$ \\
  \qquad $\dot{E}_{\mathrm{cool}}$ & $0.20$
  \enddata
  \tablecomments{See \S\ref{sec:diag-ene-budget} and
    \S\ref{sec:fid-wind-ene-budget}.}
\end{deluxetable}

\begin{figure}
  \centering
  \includegraphics[width=3.0in, keepaspectratio]
  {\figdir/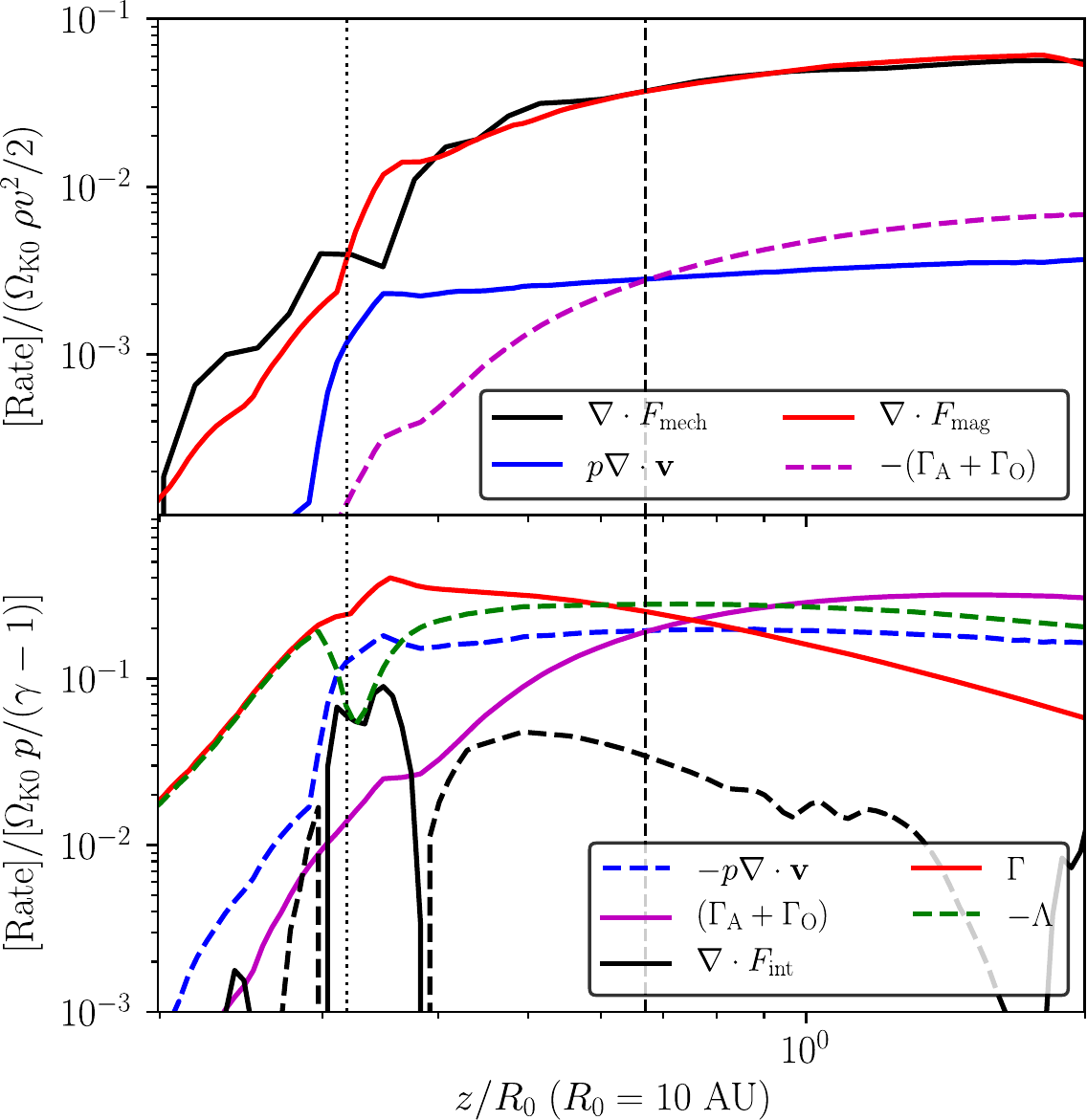}
  \caption{Local energy balance
    (eq.~\ref{eq:ene-budget-local}) along the field line
    anchored at $R_0 = 10~\au$. {\it Upper panel}:
    $\nabla \cdot F_{\mathrm{mech}}$ and associated
    quantites, normalized by $(\Omega_{\K 0} \rho v^2/2)$
    ($\Omega_{\K0}$ is the Keplerian angular velocity at
    $R_0$). {\it Lower panel}:
    $\nabla \cdot F_{\mathrm{int}}$ and associated
    quantites, normalized by $[\Omega_{\K 0}p/(\gamma-1)]$.
    Different components are distinguished by colors; line
    shape indicates the sign (solid: positive; dashed:
    negative).  Vertical dotted and dashed lines in black
    indicate the wind base and the ploidal Alfv\'enic point,
    respectively. }
  \label{fig:fiducial_energy}
\end{figure}

Analysis on the energy budget combines Figure
\ref{fig:fiducial_energy} and
Table~\ref{table:fid-ene-budget} for the fiducial model. We 
define the ``wind region'' as $\{z>z_\wb\} \cap
\{1.5<(r/\au)<50\}$; the $1<(r/\au)<1.5$ zone is excluded
due to numerical damping near the inner boundary 
(\S\ref{sec:fid-init-bound}).

The upper panel of Figure~\ref{fig:fiducial_energy} clearly
indicates that the mechanical power of the wind comes
predominantly from the magnetic field. This is confirmed by
Table~\ref{table:fid-ene-budget} with a significant negative
$\dot{E}_{\mathrm{stress}}$, indicating that the wind is
driven by the work that magnetic stress does on the wind
base, ultimately at the expense of the orbital energy of the
disk. Another important source of power is the toroidal
motion of the gas ($\dot{E}_{\mathrm{k},\phi} < 0$), which
is converted to the other forms of mechanical power inside
the wind. The majority of power injected is expended in
overcoming the negative gravitational energy of the gas;
the poloidal kinetic energy consumes the second
most of the injected power.

We notice that hard-photon heating is rather
inefficient. Out of the total hard-photon luminosity
radiated by the central star
$L_\tot\simeq 5.2\times 10^{31}~\erg~\s^{-1}$, only
$L_\mathrm{abs}\simeq 6.2\times 10^{30}~\erg~\s^{-1}$ is
actually absorbed. Eventually, less than $2~\%$ of
$L_\mathrm{abs}$ is converted to heat: most of rest is
consumed by photodissociation, photoionization and
photoelectric processes, and for maintaining $T_\dust$ (see
also appendices of WG17). However, radiative heating does
play an important role in affecting the internal energy
density near the wind base, as we can observe in the lower
panel of Figure~\ref{fig:fiducial_energy}. At higher
altitudes (roughly above the poloidal Alfv\'enic point),
heating due to non-ideal MHD dissipation, especially by
ambipolar diffusion. In total, ambipolar diffusion converts
magnetic and kinetic energy into heat at
$0.56\times 10^{30}~\erg~\s^{-1} >
|\dot{E}_\mathrm{heat}|$. Further discussions of ambipolar
heating is postponed to \S\ref{sec:pars-res-microphys}.

\begin{figure}
  \centering
  \includegraphics[width=3.0in, keepaspectratio]
  {\figdir/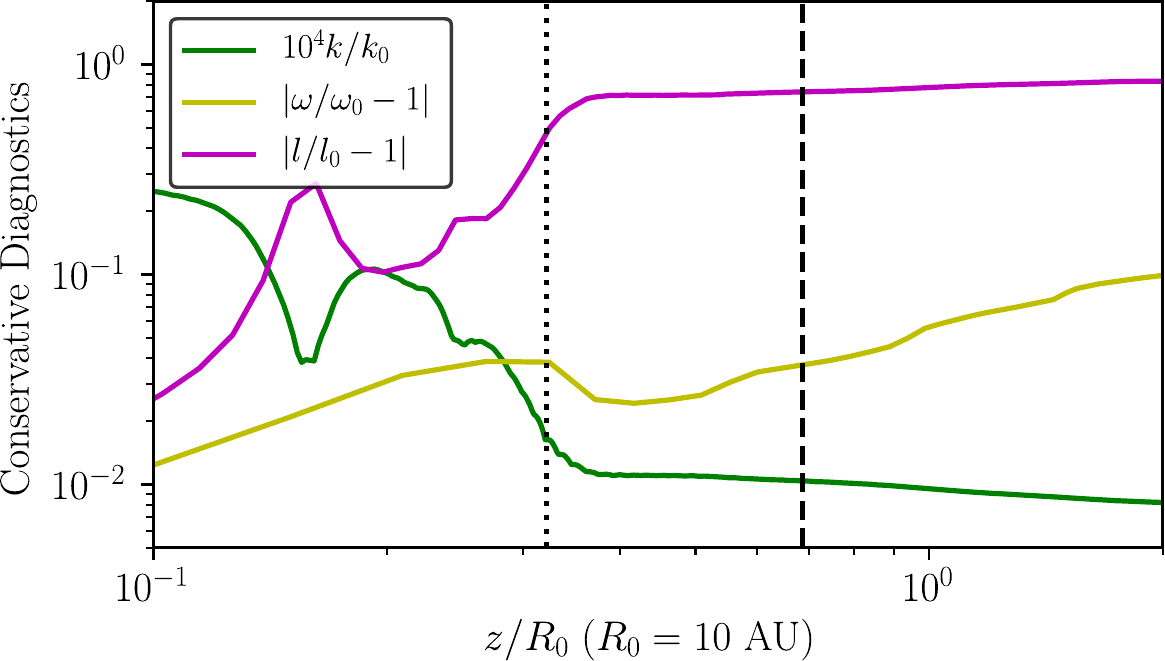}
  \caption{Like Figure~\ref{fig:fiducial_force} but for the
    wind ``constants'' $(k,\omega,l)$ defined in
    eq.~\eqref{eq:hydro-mag-diag}.}
  \label{fig:fiducial_cons_diag}
\end{figure}

\subsubsection{Magnetic field profiles}
\label{sec:fid-morph-mag}

The poloidal field lines of the fiducial model are almost
vertical near the midplane due to large diffusivities
there. Although the mid-plane diffusivity is not captured
very accurately, this behavior is largely consistent with
B17. Above the midplane in the wind, stronger field-fluid
coupling bends the poloidal field lines radially outwards.
The toroidal component $B_\phi$ dominates at almost all
radii and altitudes. This component should change sign
across the equatorial plane due to our reflecting boundary
conditions at the equatorial plane.

\subsubsection{Angular momentum budget and accretion}
\label{sec:fid-ang-mom}

\begin{figure}
  \centering
  \includegraphics[width=3.0in, keepaspectratio]
  {\figdir/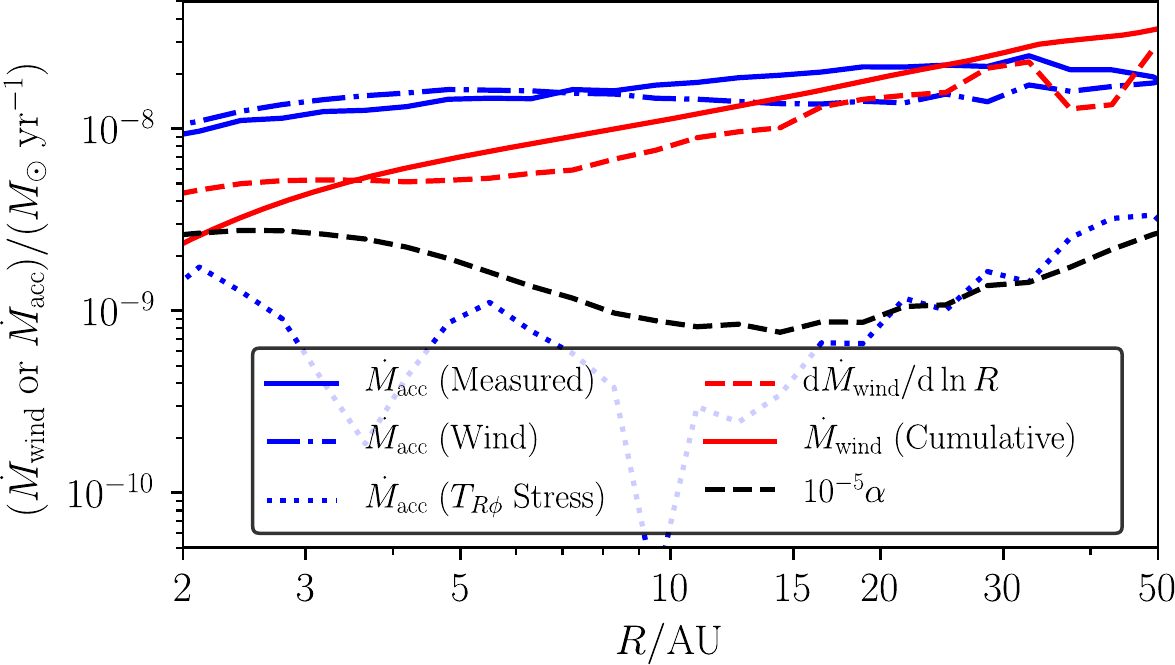}
  \caption{Radial profiles of mass dispersal rates in the
    fiducial model (Model 0). Accretion rates contributed by
    wind and $T_{R\phi}$ Maxwell stress are estimated by
    eq. \eqref{eq:acc-stress}. The Shakura-Sunyaev $\alpha$
    parameter for radial angular momentum transport (
    eq.\ref{eq:def-alpha}) is also presented.  }
  \label{fig:fiducial_acc_wind}
\end{figure}

In Figure~\ref{fig:fiducial_acc_wind} compares the accretion
rates driven by different angular momentum transport
mechanisms (eq. \ref{eq:acc-stress}). Clearly, vertical
transport of angular momentum radial transport by a
significant factor. We find that the local wind mass loss
rate is always comparable to the accretion rate, as already
discussed in \S\ref{sec:fid-wind-launch}.  The radial
Maxwell stress can be characterized by the equivalent
$\alpha$ (eq. \ref{eq:def-alpha}).  Over the entire radial
range of the fiducial simulation,
$\alpha \lesssim 3\times 10^{-4}$.

At small radii ($R\lesssim 20~\au$), we find
$R_\A/R_\wb \simeq 1.4$ and $\xi \simeq 0.5$. The lever arm
decreases to $R_\A/R_\wb\sim 1.15\pm 0.02$ at larger radii,
corresponding to bigger $\xi \simeq 1$.  These trends can be
seen in Figure~\ref{fig:fiducial_acc_wind}.  Thus while the
magnetized wind produces a reasonable accretion rate---by
construction---it does so rather inefficiently, with an even
larger outflow rate.  We expect that a larger ejection index
and a more nearly magnetocentrifugal (rather than
magnetothermal) wind could be obtained by increasing the
magnetization ($\beta_0^{-1}$).  But Model~7 (\S5) suggests
that to maintain accretion rates comparable to what is
observed, the surface density of the disk would then have to
be reduced (e.g., \citealt{2017ApJ...835...59W}).

\begin{figure}
  \centering
  \includegraphics[width=3.0in, keepaspectratio]
  {\figdir/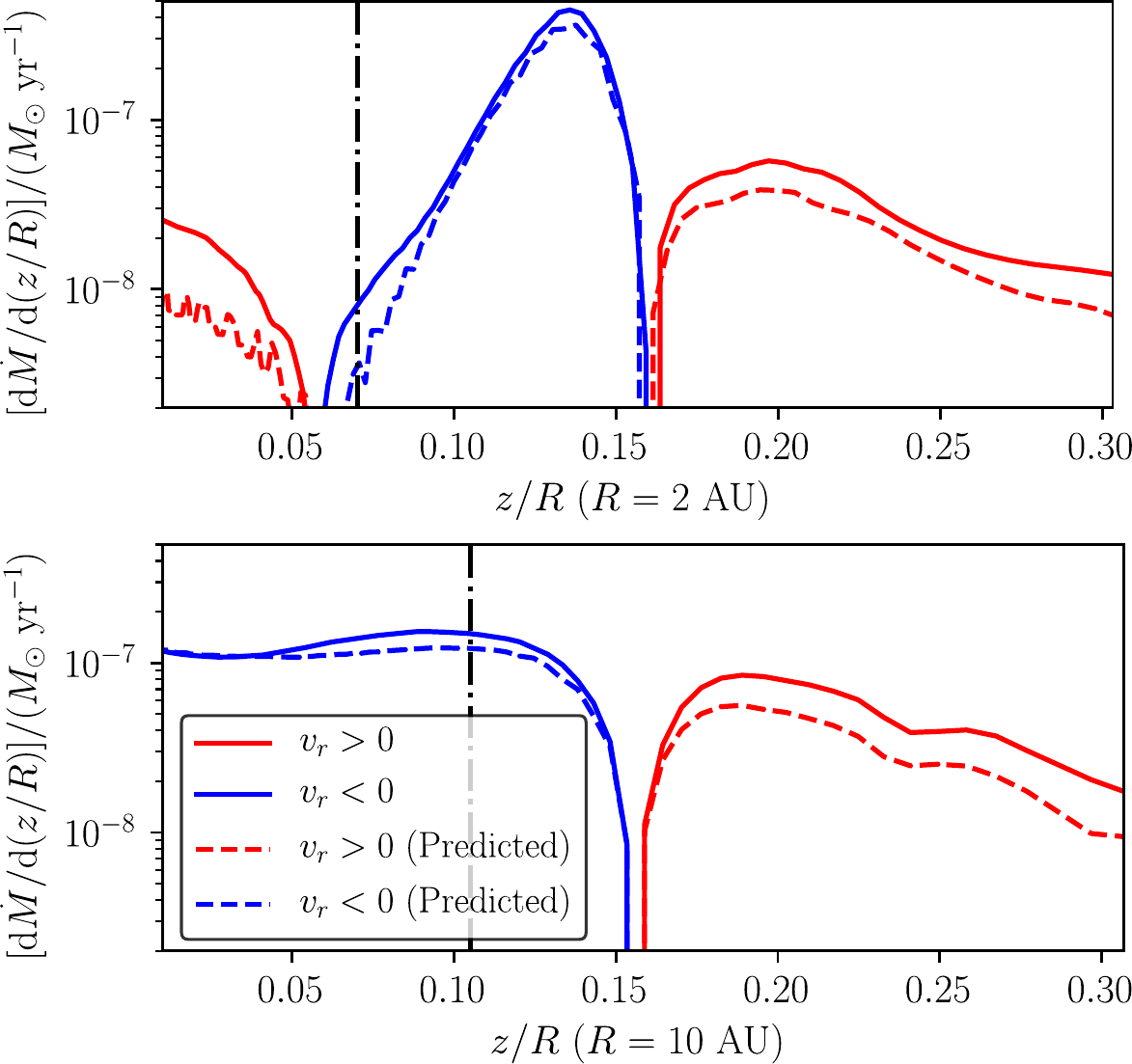}
  \caption{Vertical profiles of local mass flux [in
    $M_\odot~\yr^{-1}$ per $(z/R)$] at $R=2~\au$ and
    $R=10~\au$ for Model~0 plotted below the wind base. Blue
    and red colors indicate accretion and decretion/outflow,
    respectively. Solid lines are measured directly from the
    simulation, while dashed curves are values predicted by
    taking vertical gradients of $T_{z\phi}$ magnetic
    stress. Vertical dash-dotted line in black indicates
    $h_\mid$, the gaussian scale height at the midplane.}
  \label{fig:fiducial_radial_flow}
\end{figure}

Figure~\ref{fig:fiducial_radial_flow} exhibits vertical
profiles of the radial mass flux at two representative
cylindrical radii, $R=2~\au$ and $10~\au$.  The accreting
layer is centered at the midplane at $R=10~\au$, but at
$|z|\simeq 1.5 h_\mid$ at $R=2~\au$, evidently because the
midplane is too weakly ionized and poorly coupled in the
latter case. The dashed curves in the figure indicate radial
mass-flux profiles predicted by taking the vertical gradient
of $T_{z\phi}$ stress. These predictions are consistent with
with direct measurements of the flux, verifying that
vertical transport of angular momentum (and hence wind
launching) mediated by the $T_{z\phi}$ stress is indeed the
predominant mechanism that drives disk accretion.  There is
a layer of radial outflow above the accretion layer and
below the wind base. This outflow, like the accretion
inflow, is driven by the $B_\phi B_z$ magnetic stress, but
with the opposite sign of the vertical gradient as
$\d B_\phi /\d z$ changes its sign.  Such outflow is always
subsonic and sub-Alfv\'enic; nor does the flow actually join
the wind. In order to distinguish it from the wind, we
describe this ``decretion layer'' in what follows.  The
total outward mass flux in the decretion layer is
$\lesssim 15\%$ of the accretion. This decretion layer is a
generic feature that is also present below the wind base in
the Hall-free simulations of B17. However, as discussed in
B17, the flow structure in wind-driven accretion disks
depends on the gradient of $B_\phi$, which is sensitive to
the diffusivity profiles within the disk.  In this work, our
diffusivity profile is not necessarily realistic near the
midplane (\S\ref{sec:method-non-ideal}), thus we generally
view the flow structures there with caution.

\subsection{Thermochemistry and Radiation}
\label{sec:fid-chem-rad}


The interaction between magnetic fields and fluids is
determined by radiation and thermochemical processes. In
this subsection, we analyze the microphysics for three
different layers: the midplane (\S\ref{sec:fid-mid-plane}),
the wind-launching region (\S\ref{sec:fid-wind-launch}), and
the upper wind (\S\ref{sec:fid-wind}).

\subsubsection{Midplane region}
\label{sec:fid-mid-plane} 

The dust and gas temperatures near the midplane are
maintained somewhat artificially by the prescriptions
described in and around
eqs.~\eqref{eq:dust-temp}-\eqref{eq:prof-T_ah}.
Nevertheless, as this region is largely shielded from hard
photons and poorly coupled to the magnetic field, it would
not be appreciably affected by
photoionization/photodissociation heating or non-ideal MHD
heating in any case.  Its thermal state should properly be
regulated by the diffuse infrared radiation field, and by
thermal accomodation between dust and gas.

Though unimportant for the temperature near the midplane,
the hard photons are important for the ionization
there. Grains are the dominant charge carriers.  At small
radii ($R\lesssim 2~\au$), the ionization is very weak the
vertical column above is too massive for diffuse sources of
ionization (scattered X-ray photons and cosmic rays) to
penetrate.  The Elsasser numbers are therefore tiny,
$\am \ll \Lambda_\mathrm{Ohmic} \lesssim 10^{-4}$ at
$R=2~\au$. Eased penetration at relatively larger radii
raises the level of ionization, and the Elsasser numbers
rise accordingly: on the midplane at $10~\au$,
$\Lambda_\mathrm{Ohmic}\sim 10^{0}$ and
$\am\gtrsim 10^{-2}$.  The radial variation of midplane
ionization also leads to the change of magnetic field
morphology, as we discussed in \S\ref{sec:fid-morph-mag}.

\subsubsection{Wind-launching region}
\label{sec:fid-wind-launching-region}

At higher altitudes, the transition from a poloidally static
midplane layer to an appreciable outflow takes place near
the wind base at $3 h_\mid \lesssim z \lesssim 4 h_\mid$.

The gas density is is lower by a factor $\sim 10^{-4}$ to
$10^{-5}$ than at the midplane. Penetration is eased for
cosmic rays and scattered X-rays, and especially for high
energy photons propagating directly from the
protostar. Thanks to these processes, the gas temperature
rises to about twice that of the local midplane. The hard
photons also yield a considerably higher level of
ionization. As the absolute abundance of charge increases to
$\sim 10^{0\pm 1}~\cm^{-3}$ ( fractional ionization
$\sim 10^{-7\pm 1}$), the ambipolar Elsasser number $\am$
becomes $\sim O(1)$.  The dominant charge carriers are
\pos{S} and \pos{Si}, while carbon is almost neutral since
photoionization of \chem{CO} and C are more susceptible to
cross-/self-shielding effects, especially by \chem{H_2}.

\subsubsection{Wind region}
\label{sec:fid-wind}

The physics is relatively straightforward inside the
outflow.  The gas continues to be accelerated by the
magnetic pressure gradient all the way through the
Alfv\'enic point.  The low gas density and column density
allows more ionizing photon to get in, hence the ambipolar
Elsasser number reaches $30 \lesssim \am \lesssim 10^2$.
Even such an $\am$ allows some drift between charged and
neutral particles, therefore field lines and streamlines do
not perfectly coincide, and there is significant ambipolar
heating.

The temperature of the wind is determined by the balance
between ambipolar diffusion heating (the leading heating
mechanism), and adiabatic expansion (for the outflow
launched from $R\lesssim 5~\au$) and/or atomic/molecular
cooling (for the outflow launched by the $R\gtrsim
5~\au$). Radiative heating is relatively ineffective because
of the small X-ray cross section, strong cross-shielding of
LW photons, and depletion of molecular species susceptible
to dissociation by soft FUV. At relatively large radii, the
gas temperature drops slightly as fluids move outwards,
since the sum of adiabatic and radiative cooling mechanisms
exceeds non-ideal MHD heating (see
Figures~\ref{fig:fiducial_vert}, \ref{fig:fiducial_energy}).

\begin{figure}
  \centering
  \includegraphics[width=3.0in, keepaspectratio]
  {\figdir/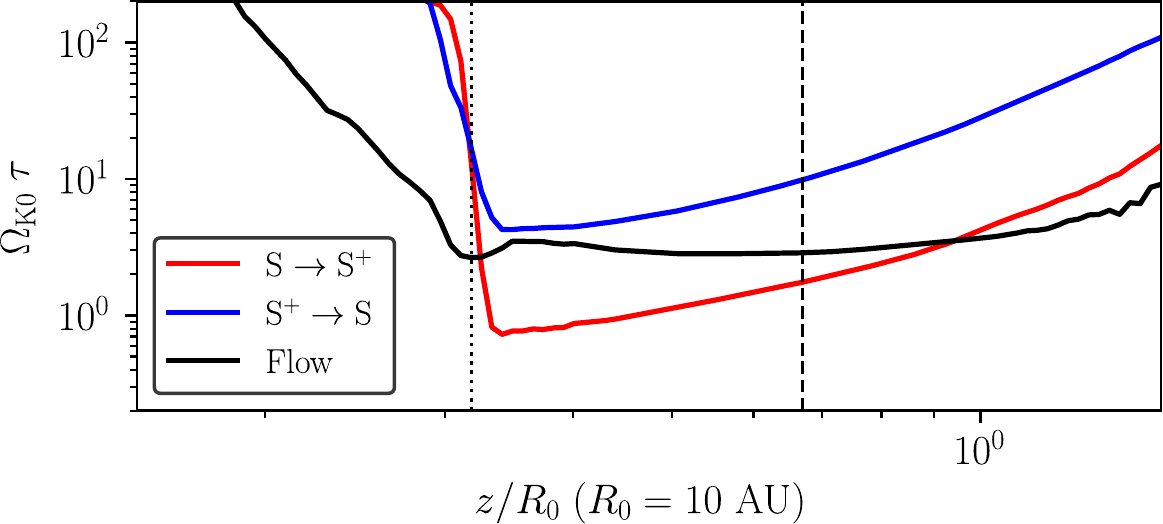}
  \caption{Timescales of hydrodynamic flow, and radiative
    ionization and recombination of sulfur along the field
    line anchored at $R_0=10~\au$
    (eq.~\ref{eq:def-flow-time}). All timescales are
    normalized by $\Omega_{\K 0}$, the Keplerian angular
    velocity at $R_0=10~\au$. }
  \label{fig:fiducial_tau_s_flow}
\end{figure}

\begin{figure*}
  \centering
  \includegraphics[width=7.0in, keepaspectratio]
  {\figdir/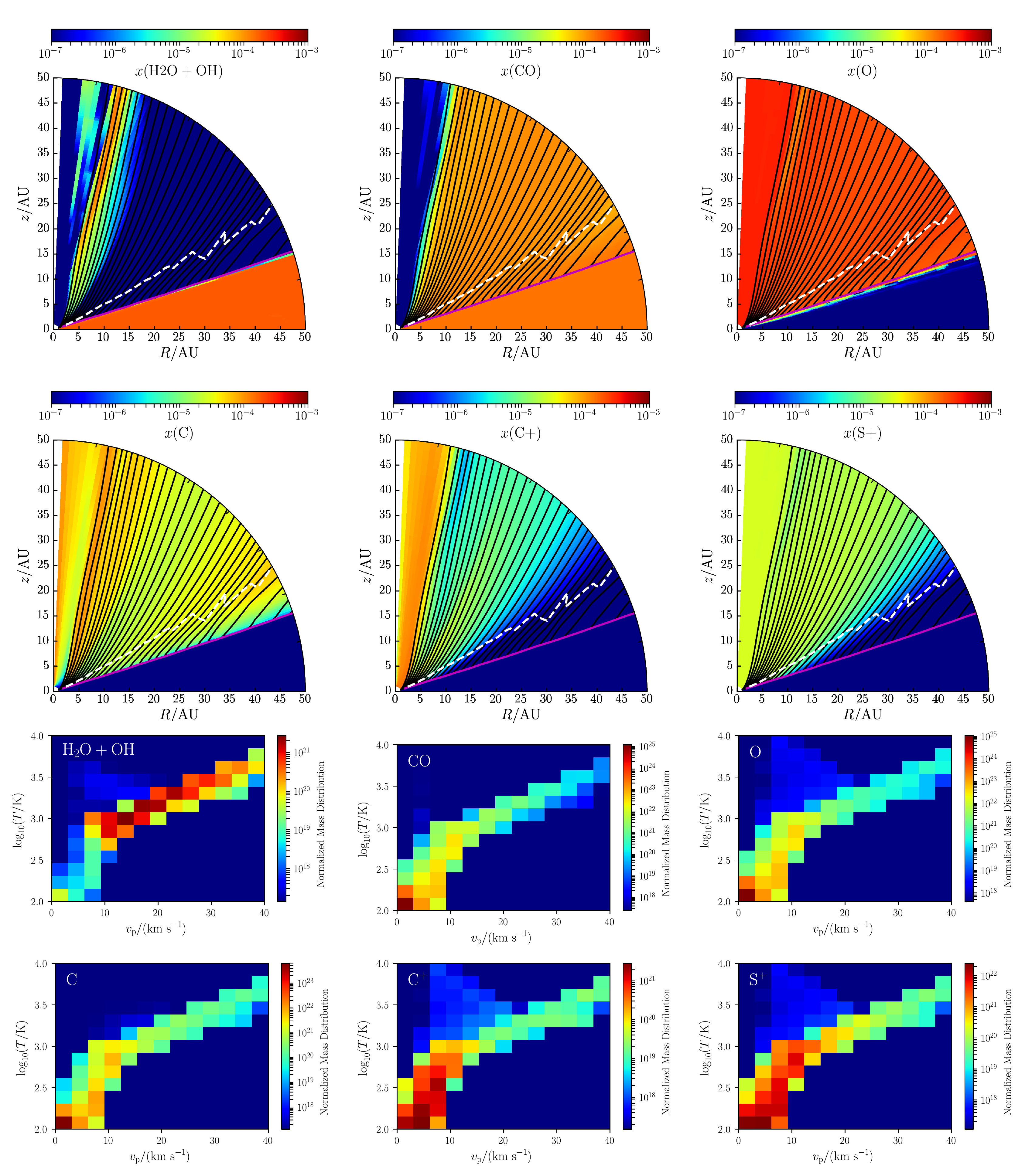}
  \caption{ Distribution of key species in Model~0:
    \chem{H_2O}/OH, CO, O (atomic oxygen), C (atomic
    carbon), \pos{C}, and \pos{S}. Top two rows show their
    spatial distribution, while bottom two rows present
    their distributions in the two-dimensional space of
    poloidal speed $v_\p$ and logarithm of temperature.  The
    normalized mass distribution represents
    $\d^2m / [\d\log_{10}(T/\K)\d(v_\p/\kms)]$.  Stream
    lines, wind bases, and Alfv\'enic surfaces are indicated
    in the top two rows as in
    Figure~\ref{fig:fiducial_slice}. }
  \label{fig:fiducial_phase_sp}
\end{figure*}

Figure~\ref{fig:fiducial_phase_sp} plots the distributions
of several key chemical species in velocity and
temperature. We notice a prominent tail of neutral
atomic/molecular species inside the wind, located at
relatively high temperature ($T\gtrsim 10^3~\K$) and
intermediate poloidal velocity ($v_\p\sim 15~\km~\s^{-1}$).
In Table~\ref{table:fid-mass-loss-sp}, we present the
mass-loss rates of key species. The magneto-thermal wind is
predominantly molecular, thanks to its high density in
comparison to the unmagnetized EUV-driven winds studied by
WG17. Note that abundances of those species
cannot be correctly calculated if we assume local
thermochemical equilibrium instead of evolving
thermochemical networks in real-time. As an example, we
compare the flow timescale to the timescales
of radiative ionization and recombination of sulfur along a
field line. We define these as,
\begin{equation}
  \label{eq:def-flow-time}
  \begin{split}
    & \tau_\mathrm{flow} \equiv \left| v_\p\ \dfrac{\d \ln
        \rho}{\d s} \right|^{-1}\ ,\ 
    \tau_{\chem{S}\rightarrow \pos{S}} \equiv
    \dfrac{n(\chem{S})}
    {k(h\nu + \chem{S}\rightarrow e^-+ \pos{S})}\ ,
    \\
    & \tau_{\pos{S}\rightarrow \chem{S}} \equiv
    \dfrac{n(\pos{S})}
    {k(\pos{S} + e^-\rightarrow \chem{S})}\ ,  
  \end{split}
\end{equation}
where the derivative in $\tau_\mathrm{flow}$ is taken with
respect to arc length $s$ along the field line, and $k$
denote the reaction rates. In
Figure~\ref{fig:fiducial_tau_s_flow}, at all altitudes above
the wind base, $\tau_\mathrm{flow}$ is comparable to the
timescales of ionization and recombination. 

\begin{deluxetable}{lll}
  \tablecolumns{3} 
  \tabletypesize{\scriptsize}
  \setlength{\tabcolsep}{20pt}    
  \tablecaption{Mass-loss rates of several species in
    Model 0.
    \label{table:fid-mass-loss-sp}
  }
  \tablehead{
    \colhead{Species} &
    \colhead{$\dot{M}$} &
    \colhead{$\dot{M}/\dot{M}_\wind$} 
    \\
    \colhead{} &
    \colhead{($M_\odot~\yr^{-1}$)} &
    \colhead{}
  }
  \startdata
  \chem{H_2} & $1.91 \times 10^{-8}$ & $6.97 \times 10^{-1}$
  \\ 
  \chem{H} & $5.64 \times 10^{-10}$ & $2.06 \times 10^{-2}$
  \\ 
  \pos{H} & $2.56 \times 10^{-14}$ & $9.34 \times 10^{-7}$ \\
  \chem{He} & $7.81 \times 10^{-9}$ & $2.85 \times 10^{-1}$
  \\ 
  \pos{He} & $1.90 \times 10^{-14}$ & $6.93 \times 10^{-7}$
  \\ 
  \chem{H_2O} & $9.01 \times 10^{-13}$ & $3.29 \times
  10^{-5}$ \\ 
  \chem{OH} & $2.21 \times 10^{-12}$ & $8.07 \times 10^{-5}$
  \\ 
  \chem{CO} & $5.38 \times 10^{-11}$ & $1.96 \times 10^{-3}$
  \\ 
  \chem{O} & $6.67 \times 10^{-11}$ & $2.43 \times 10^{-3}$
  \\ 
  \pos{O} & $7.80 \times 10^{-17}$ & $2.84 \times 10^{-9}$ \\
  \chem{C} & $9.02 \times 10^{-12}$ & $3.29 \times 10^{-4}$
  \\ 
  \pos{C} & $7.00 \times 10^{-13}$ & $2.55 \times 10^{-5}$ \\
  \chem{S} & $1.07 \times 10^{-11}$ & $3.89 \times 10^{-4}$
  \\ 
  \pos{S} & $6.84 \times 10^{-12}$ & $2.49 \times 10^{-4}$ 
  \enddata
\end{deluxetable}

We attribute the co-existence of CO and \pos{C} in the wind
to shielding effects: in the LW band, CO photodissociation
is more sensitive to self-shielding and especially to
cross-shielding by \chem{H_2}, compared to photoionization
of CO and C. As a result, although LW photons penetrate
deeply near the wind base, CO still exists there.

\chem{H_2O} and OH molecules are mainly dissociated by the
soft FUV continuum, hence they are not sensitive to
shielding effects and do not survive in significant numbers
in the wind. Instead, oxygen exists mostly in atomic form,
with poloidal velocity
($v_{\mathrm{p}}\sim 15~\km~\s^{-1} $). At high latitudes,
however, where $T\sim 10^3~\K$ and
$n(\chem{H_2})\sim 10^{6-7}~\cm^{-3}$, result in the
re-formation rates of \chem{H_2O} and \chem{OH} are
competitive with photodissociation (see the upper-left and
upper-middle panels in Figure~\ref{fig:fiducial_slice}, and
the upper left panel in Figure~\ref{fig:fiducial_phase_sp}).
For example, at $r=5~\au$ and
$0.2 \lesssim \theta \lesssim 0.4$ in our fiducial model,
the major destruction mechanism of \chem{OH} molecules is
photodissociation by soft FUV photons, the rate per
\chem{OH} molecule being $\xi\simeq 3\times 10^{-5}~\s^{-1}$
(absorption ignored).  At $T\simeq 10^3~\K$ and
$n(\chem{H_2})\simeq 10^7~\cm^{-3}$, the rate at which an
oxygen atom re-forms \chem{OH} is approximately (according
to the \verb|UMIST| database; see \citealt{UMIST2013})
$3\times 10^{-6}~\s^{-1}$, via the most important reaction
channel
$\chem{O} + \chem{H_2} \rightarrow \chem{\chem{OH}} +
\chem{H}$.  As a result, $\sim 10~\%$ of the oxygen resides
in \chem{OH} molecules.  These re-formed
\chem{H_2O}/\chem{OH} molecules mainly appear at very high
latitudes and intermediate radial velocity
($v_r\sim 10-40~\kms$), and of course at $T\sim 10^3\,\K$.
The distribution of these molecules could be checked
observationally and used to constrain wind models.

\section{Parameter Study}
\label{sec:res-param-space}

To explore the sensitivity of our results to our input
parameters, we performed a number of additional simulations,
each differing from the fiducial run in one parameter.
These simulations are described in
\S\ref{sec:pars-models-setup} and summarized in
Table~\ref{table:var-model}.

\begin{deluxetable}{lccc}[!t]
  \tablecolumns{5} 
  \tabletypesize{\scriptsize}
  \tablecaption{Various models for parameter study
    \label{table:var-model}
  }
  \setlength{\tabcolsep}{3pt}
  \tablehead{
    \colhead{Model} &
    \colhead{$\dot{M}_\acc(10~\au)$} &
    \colhead{$\dot{M}_\wind$} &
    \colhead{$\mean{v_r}$}
    \\
    \colhead{} &
    \colhead{$(10^{-8}~M_\odot/\yr)$} &
    \colhead{$(10^{-8}~M_\odot/\yr)$} &
    \colhead{$(\kms)$}
    \\
    \colhead{(1)} &
    \colhead{(2)} &
    \colhead{(3)} &
    \colhead{(4)}
  }
  \startdata
  0 (Fiducial) & $1.7\pm 0.3$ & $2.9\pm 0.2$ & $4.6\pm 0.4$
  \\
  $0^-$ (Convergence test) & $1.9\pm 0.5$ & $2.7\pm 0.4$
  & $4.0\pm 0.3$  \\
  $0^\gamma$ ($\gamma = 1.4$) & $1.3\pm 0.3$ & $2.7\pm 0.2$
  & $4.4\pm 0.6$ \\
  1 (Ambipolar heating off) & $1.7 \pm 0.3$
  & $2.8\pm 0.3$  & $4.7\pm 0.3$ \\
  2 ($0.1\times $ X-ray) & $1.2\pm 0.4$ & $2.7 \pm 0.3$ &
  $6.0 \pm 0.3$ \\
  3 ($0.1\times $ Soft FUV) & $1.7\pm 0.4$ & $2.4\pm 0.3$ &
  $5.5 \pm 0.7$ \\
  4 ($0.1\times $ LW) & $1.6 \pm 0.3$ & $2.6\pm 0.2$ &
  $5.0\pm 0.5$ \\
  5 ($0.1\times $ All hard photons) & $1.1 \pm 0.5$ &
  $1.4\pm 0.4$ & $7.5\pm 1.2$ \\  
  6 ($\Phi_\euv = 10^{41.7}~\s^{-1}$) & $1.5\pm 0.4$ &
  $1.6\pm 0.1$ & $8.5\pm 0.6$ \\
  7 ($\beta_0 = 10^4$) & $17.1\pm 1.5$ & $16.6\pm 3.3$ &
  $6.7\pm 0.7$ \\
  8 ($0.1\times $ Dust grains$^*$) & $7.4 \pm 0.8$
  & $5.1\pm 1.5$ & $3.4\pm 4.5$ \\
  \enddata
  \tablecomments{
    (1) Model identifier and description (parameter by which
    model differs from fiducial). 
    (2) Accretion rate measured at $R=10~\au$.
    (3) Wind mass-loss rate.
    (4) Mean outflow velocity (weighted by mass flux).
    Measurements are averaged over the final $50~\yr$ of
    each simulation. Quoted errors are twice the r.m.s. time
    variation. 
    \\
    $*$: This model exhibits instabilities; see
    \S\ref{sec:pars-res-dust}}
\end{deluxetable}

\subsection{Parameter variations}
\label{sec:pars-models-setup}

Model $0^-$ tests the convergence of our simulations,
repeating the fiducial run on a $192\times 64$ grid, i.e. 2
times coarser in both latitudinal and radial zones.  Model
$0^\gamma$ recalculates the fiducial model with
$\gamma = 1.4$ rather than $\gamma=5/3$ in the hydrodynamic
solver.

Our analyses in \S\ref{sec:fid-chem-rad} show that heating
by non-ideal MHD effects could be important in at least some
regions of the wind-launching system.  Model 1 further
explores this by turning off such heating processes.

According to WG17, the the spectral-energy distribution of
the high-energy radiation is crucial for shaping
photoevaporative outflows in the absence of magnetic
fields. One major concern is therefore how different bands
of radiation affect MHD outflows. Specifically, the fiducial
model omits EUV radiation for better comparison with B17.
This may not be realistic, as EUV may still reach the top of
the wind---i.e., the parts nearest the axis, which have the
lowest radial column density.  Models 2 through 6 examine
the importance of radiation, especially EUV, by varying the
luminosities in each energy band.

The magnetic field plays a central role in transporting
angular momentum and hence disk dispersal. Model 7 examines
the impact of increased magnetization by setting midplane
plasma $\beta_0=10^4$.

Models 8 varies the abundance of dust grains.
Arguments in \citet{2006A&A...459..545G,
  2008ARA&A..46..289T, 2016A&A...586A.103W} suggest that our
PAH abundance is roughly 4 to 10 times the ISM standard.
This is deliberate, because we use PAH as a proxy for all
dust grains.  Model 8, however, reduces the abundance of PAH
by a factor of 10.

\subsection{Numerical effects}
\label{sec:pars-verify-fid}

The grid resolution and the restriction to a constant
adiabatic index in the hydrodynamic solver are numerical or
algorithmic limitations rather than astronomical or physical
uncertainties.  Model $0^-$ shows almost identical MHD
profiles and consistent wind mass loss rates to those of the
fiducial model, thus verifying the convergence of the
latter.

According to \citet{2016ApJ...818..152B}, wind mass loss
rates should be relatively insensitive to the adiabatic
index.  Model $0^\gamma$ appears to confirm this, although
the accretion rate decreases by $\sim 20\%$. Fully detailed
and consistent treatment will require nontrivial technical
improvements to the hydrodynamic solver that are left to
future works.

\subsection{Heating, ionization, diffusivity
  and dynamics} 
\label{sec:pars-res-microphys}

We have suggested that microphysics is important for the
launching and structure of PPD winds. Nevertheless, Models 1
through 5 are qualitatively similar to the fiducial model,
though with some quantitative differences in outflow or
accretion rate.

Despite the neglect of ambipolar heating in Model 1, the
quantitative results are indistinguishable from the fiducial
within their errorbars.  This is apparently because, as
Figure~\ref{fig:fiducial_energy} shows, non-ideal MHD
heating is important---in the sense that the associated
heating timescale is comparable to the flow timescale---
only at relatively high altitudes, where the wind is already
super-Alfv\'enic.  Nevertheless, the neglect of non-ideal
MHD heating lowers the wind temperature to $T\sim 300~\K$,
compared to $T\sim 10^3~\K$ in the fiducial model at
$0.3\lesssim \theta \lesssim 0.6$.
\citet{1993ApJ...408..115S, 2001A&A...377..589G} suggested
that ambipolar heating might be important in PPD wind
structures.  Our results suggest that, at least in these
magneto-thermal winds, such heating is more important for
the temperature of the wind than for its outflow speed or
mass-loss rate.

\begin{figure}
  \centering
  \includegraphics[width=3.0in, keepaspectratio]
  {\figdir/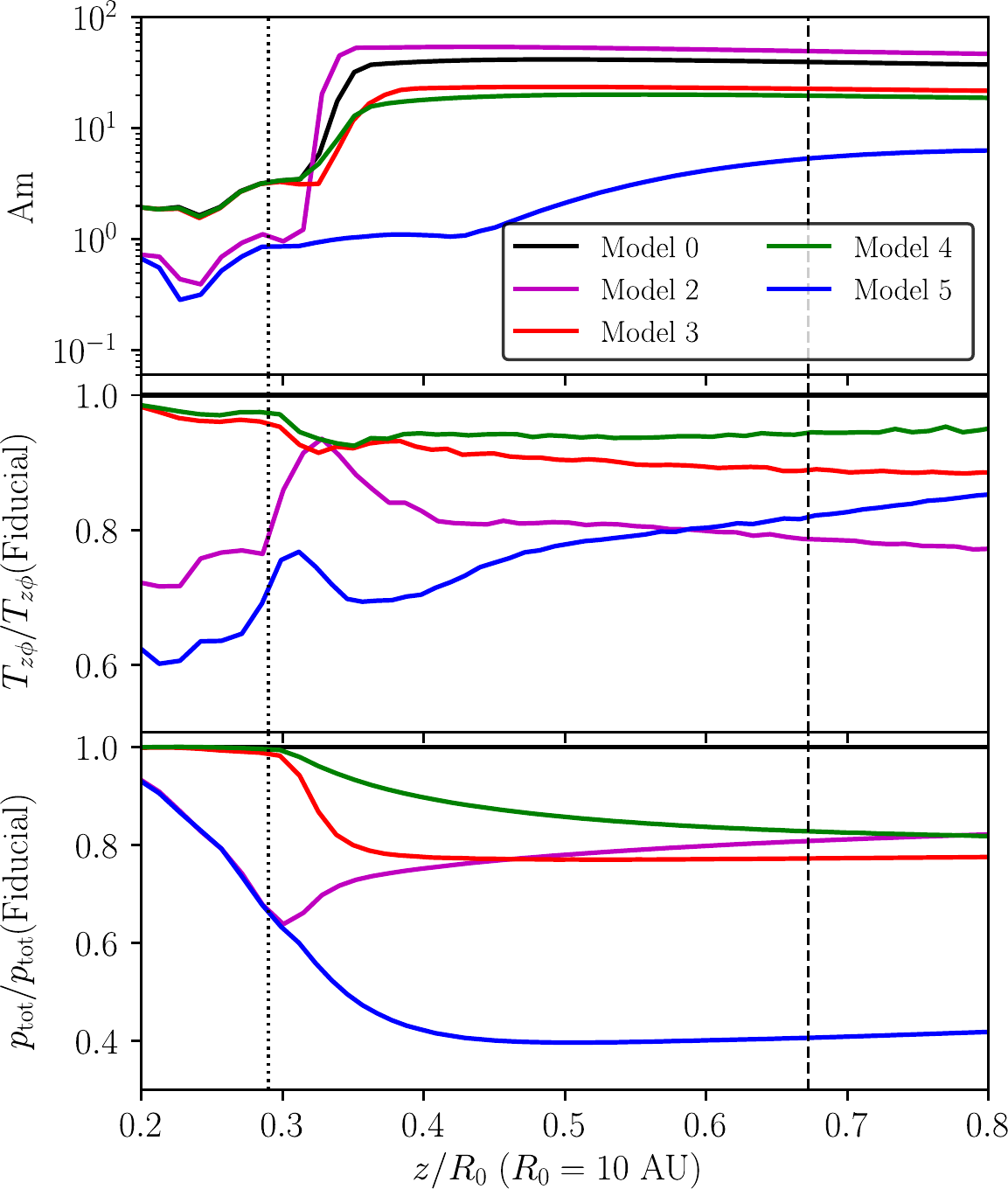}
  \caption{Ambipolar Elsasser number (top panel), $z-\phi$
    Maxwell stress $T_{z\phi}$ (middle panel) and total
    pressure $p_\tot$ (bottom panel) for Models 0, 2, 3, 4,
    5, along field lines anchored at
    $R_0=10~\au$. $T_{z\phi}$ and $p_\tot$ profiles are
    normalized to those of the fiducial model (Model
    0). Vertical dotted and dashed lines mark the wind base
    and the poloidal Alfv\'enic point of Model 0,
    respectively. }
  \label{fig:compare_am_mag}
\end{figure}

Models 2 through 5 test the response to reductions in the
various hard-photon luminosities. Reduction of the X-ray
luminosity has the greatest effect on the accretion rate
(Model 2), while reduction of the soft-FUV and LW
luminosities affect the wind mass-loss rate more than the
accretion rate.

What is the underlying reason for the decreased accretion
and mass-loss rates? In principle, $\dot{M}_\wind$ can be
reduced by reducing the heating rate at the wind
base. However, as we can observe from
Table~\ref{table:fid-ene-budget}, radiative heating takes a
small share ($\sim 5~\%$) in the total energy balance of the
wind; it can only affect wind launching weakly. Instead,
both wind and accretion are driven mainly by magnetic
fields, whose effect can be undercut by reducing the
coupling between the gas and the field (reducing $\am$).

For lower X-ray luminosity, this undercutting happens most
in and near the midplane. Scattered X-rays are the dominant
ionization mechanism immediately below the wind base, as
their absorption length is the longest among the hard-photon
bands. With an X-ray luminosity ten times less than that of
the other models, Model 2 has an $\am$ near and below the
wind base that is reduced by $\sim 2-3$ times. The weaker
coupling leads to weaker torodial fields by $\sim 20~\%$;
this reduces both the toroidal magnetic pressure that helps
to propel the outflow ($\propto B_\phi^2$), and also the
torque on the disk that drives accretion
($\propto T_{z\phi} \propto B_\phi B_z$), as we can observe
in Figure~\ref{fig:compare_am_mag}.

Soft-FUV and LW photons do not efficiently penetrate below
the wind base; their luminosities do not affect the
gas-field coupling there. Thus Models 3 and 4, with the
same X-ray luminosity as the fiducial model, do not have
appreciable difference in the wind-base $T_{z\phi}$ stress
profiles and the accretion rates. Instead, their reduced
luminosities in soft-FUV or LW result in weaker coupling
only above the wind bases, reducing toroidal
magneticpressure and hence the wind mass-loss rates.

Model 5 is fainter than the fiducial model in all hard
photon bands (excepting EUV, which both models lack
entirely). It has similar rate of accretion as Model 2 due
to lower X-ray luminosity, while the reduction in wind
mass-loss rate combines the influence.


\subsection{EUV Hybrid Wind}
\label{sec:pars-res-euv}

\begin{figure*}
  \centering
  \includegraphics[width=7.0in, keepaspectratio]
  {\figdir/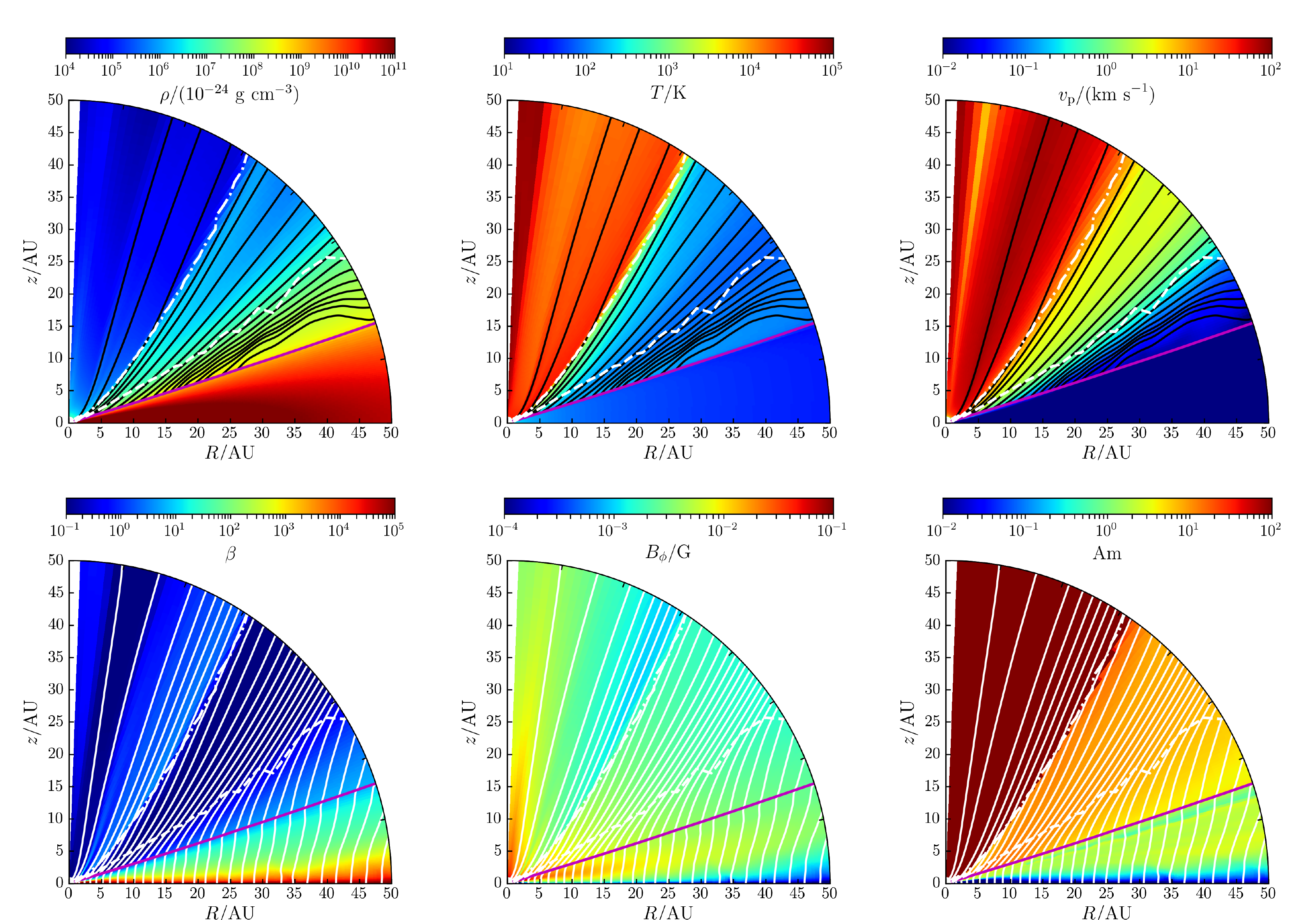}
  \caption{Same as Figure~\ref{fig:fiducial_slice}, but
    plotted for Model 6. White dash-dotted curves indicate
    the EUV front. }
  \label{fig:euv1x_slice}
\end{figure*}

\begin{figure}
  \centering
  \includegraphics[width=3.0in, keepaspectratio]
  {\figdir/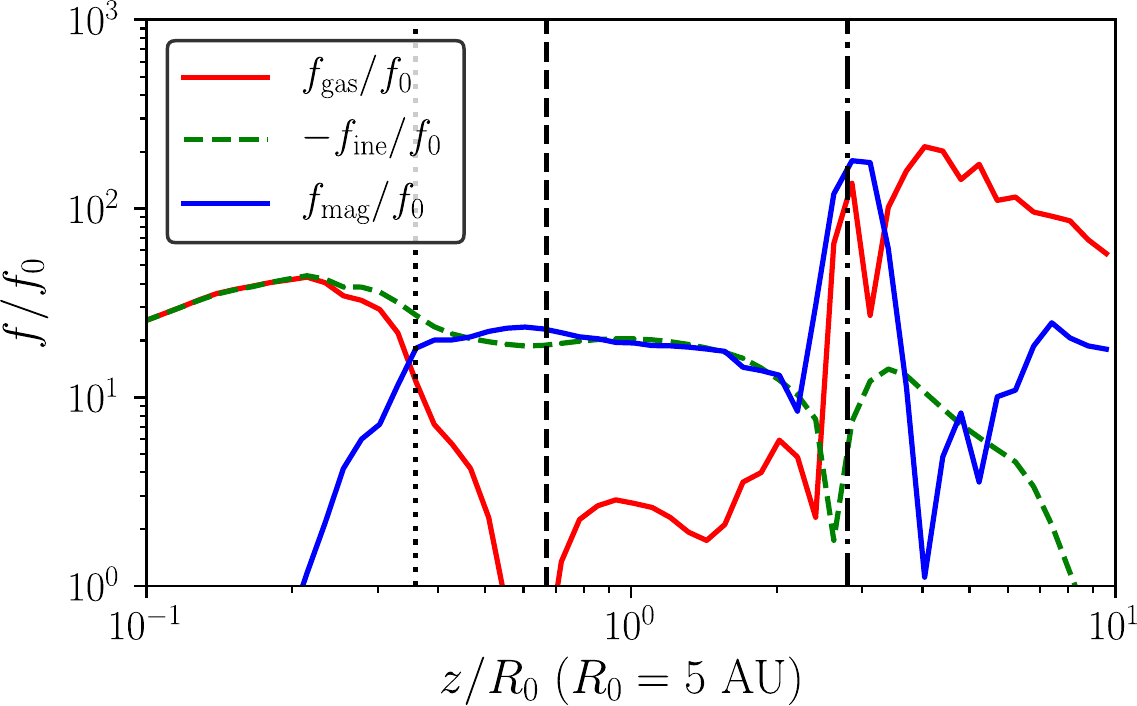}
  \caption{ Like Figure~\ref{fig:fiducial_force}, but for a
    field line anchored at $R_0=5\,\au$ in Model~5. Vertical
    dash-dotted line indicates the EUV front.}
  \label{fig:euv1x_force}
\end{figure}

\begin{figure}
  \centering
  \includegraphics[width=3.0in, keepaspectratio]
  {\figdir/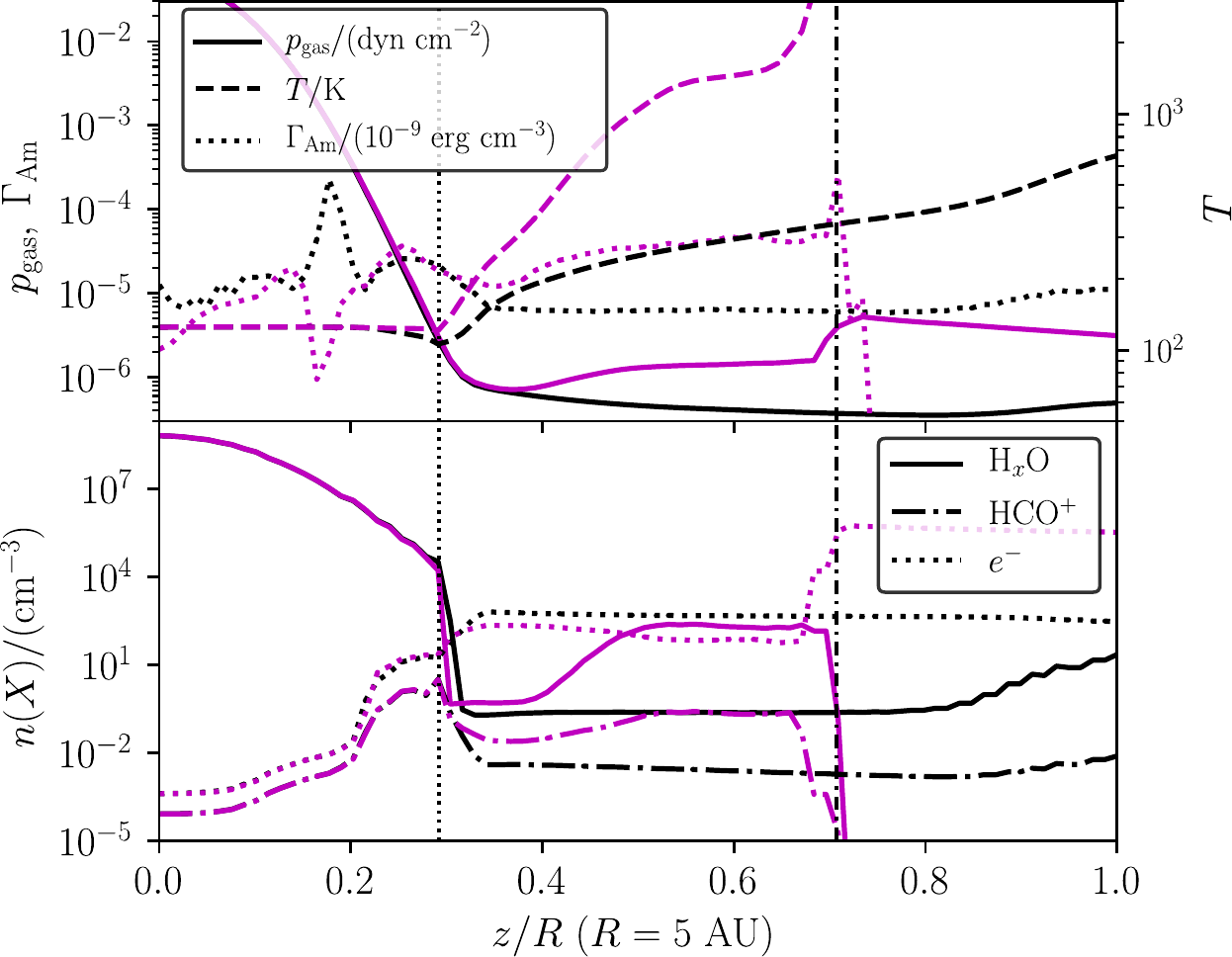}
  \caption{Comparison of vertical profiles at $R=5~\au$ for
    Models 0 and 6. {\it Upper panel}: gas pressure
    $p_\mathrm{gas}$ (left ordinate), ambipolar heating rate
    $\Gamma_\am$ (left ordinate), and temperature $T$ (right
    ordinate). {\it Lower panel}: abundances of \chem{H_xO}
    (\chem{H_2O} and OH combined; lower panel), \pos{HCO}
    and $e^-$ (lower panel). Models are distinguished by
    colors (Model 0: black; Model 6: magenta). Vertical
    dotted line marks the magneto-thermal wind bases for
    both models; vertical dash-dotted line indicates the EUV
    front. }
  \label{fig:compare_fid_euv}
\end{figure}

\begin{figure*}
  \centering
  \includegraphics[width=7.0in, keepaspectratio]
  {\figdir/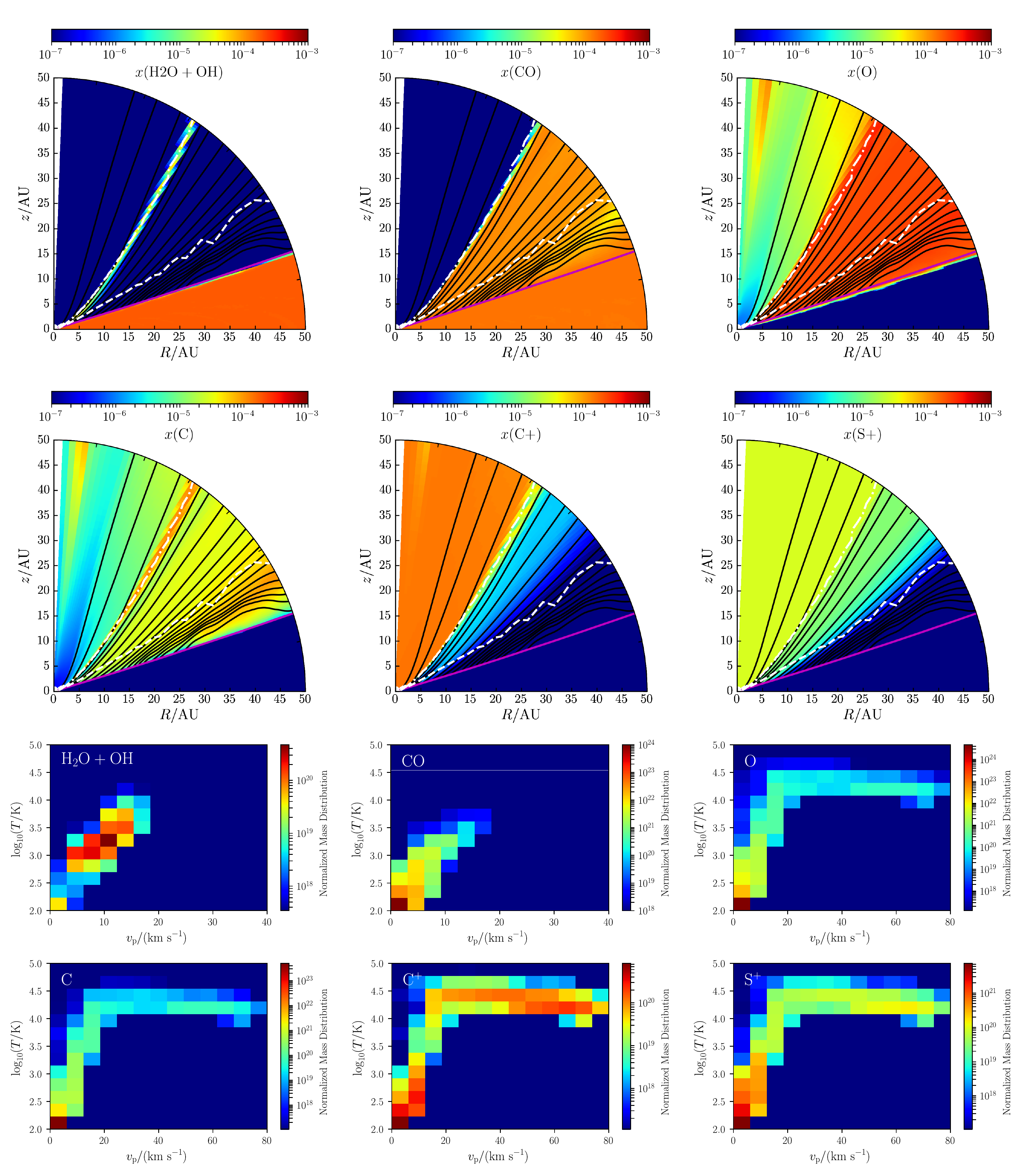}
  \caption{ Like Figure~\ref{fig:fiducial_phase_sp} but for
    Model 6. White dash-dotted curves indicate the EUV
    front. Note that the ranges of temperature and velocity
    are different.}
  \label{fig:euv1x_phase_sp}
\end{figure*}

EUV photons have much larger absorption cross sections on
hydrogen and helium than FUV and X-ray photons, and hence
can only penetrate a radial column
$N_{\mathrm{H}}\sim 10^{18}~\cm^{-2}\sim 10^5~\cm^{-3}\times
10~\au$, Thus, compared to these other high-energy bands,
the EUV heats a smaller amount of mass to much higher
temperatures.  With the addition of EUV, Model 6 adds a fast
component to the outflow that that is not present in the
fiducial model.

We present meridional plots of Model~5 in
Figure~\ref{fig:euv1x_slice}. The most prominent feature is
that the outflow is a hybrid of two components. Above a
magneto-thermal wind that is qualitatively similar to that
of non-EUV models, there is a fast EUV wind, in which the
gas temperature rises to $T\sim 2\times10^4~\K$ and poloidal
velocities reach $v_\p \sim 40-80~\kms$. This fast wind
``steals'' $\sim 0.26\times 10^{-8}~M_\odot~\yr^{-1}$ from
the magneto-thermal wind beneath.

Figure~\ref{fig:euv1x_force} illustrates the force
decomposition along the field line emanating from
$R_0=5~\au$, in which we can easily recognize the EUV front
at $z/R_0\simeq 3$. The contributions of the magnetic
pressure gradient $f_\mathrm{gas}$ and of the gas pressure
gradient $f_\mathrm{mag}$ are comparable in launching this
wind. Obviously $f_\mathrm{gas}$ is dominated by EUV heating
above the wind base.  The rapid acceleration and expansion
of the flow just above the base causes an abrupt decrease in
magnetic field strength (see also lower-middle panel of
Figure~\ref{fig:euv1x_slice}) and thus a significant
gradient of magnetic pressure, which is reflected by a peak
in $f_\mathrm{mag}$.

In the magneto-thermal wind, Model 6 has a lower mass-loss
rate than Model 0.  In Figure~\ref{fig:compare_fid_euv} we
compare these two models by showing vertical profiles at
$R=5~\au$.  Launching of the EUV wind exerts pressure on the
gas below the EUV front. The magneto-thermal wind adjusts
itself to the extra pressure by converging to a new state
that has higher temperature and similar density profile
compared to its counterpart in the fiducial model. This
higher temperature is maintained by increased ambipolar
dissipation, which in turn is a result of reduced ion
density due to faster recombination at higher temperature.
For example, the re-formation rates of OH and \chem{H_2O}
molecules (\chem{H_xO}) are faster in the magneto-thermal
wind of Model 6, which subsequently yields more molecular
ions (e.g. \pos{HCO}) by charge exchange reactions. Because
these molecular ions are very efficient eliminator of free
electric charges via dissociative recombination, coupling
between the gas and the field is weaker in the
magneto-thermal wind layer, leading to the contrast the
ambipolar Elsasser number profiles presented in
Figures~\ref{fig:fiducial_slice} versus
\ref{fig:euv1x_slice}. This effect further reduces the
magnetic pressure in Model 6, and ultimately the wind mass
loss rate.

\begin{deluxetable}{lrr}
  \tablecolumns{3} 
  \tabletypesize{\scriptsize}
  \tablewidth{500pt}
  \tablecaption{Model 6 wind energy budget.
    \label{table:euv-ene-budget}}
  \tablehead{
    Item & \colhead{Magneto-thermal wind} &
    \colhead{EUV wind}\\
    & { [$10^{30}~\erg~\s^{-1}$]} & {[$10^{30}~\erg~\s^{-1}$]}
  }
  \startdata
  $\dot{E}_{\mathrm{mech}}$ & $0.75$ & $1.97$ \\ [2pt]
  \qquad $\dot{E}_{\mathrm{k,p}}$ & $0.06$ & $1.72$ \\
  \qquad $\dot{E}_{\mathrm{k},\phi}$ & $-0.74$ & $-0.22$ \\
  \qquad $\dot{E}_{\mathrm{grav}}$ & $1.40$ & $0.41$ \\
  \qquad $\dot{E}_{p\d V}$ & $0.01$ & $0.06$ \\ [5pt]
  $\dot{E}_{\mathrm{mag}}$ & $-0.90$ & $-0.67$ \\ [2pt]
  \qquad $\dot{E}_{\mathrm{stress}}$ & $-1.14$ & $-0.28$ \\
  \qquad $\dot{E}_{\mathbf{S}'}$ & $0.27$ & $-0.33$ \\
  \qquad $\dot{E}_{\B}$ &  $-0.03$ & $-0.06$ \\[5pt]
  $\dot{E}_{\mathrm{int}}$ & $0.02$ & $0.09$ \\[5pt]
  $\dot{E}_{\mathrm{na}}$ & $0.13$ & $-1.38$ \\ [2pt]
  \qquad $\dot{E}_{\mathrm{heat}}$ & $-0.11$ & $-2.02$ \\
  \qquad $\dot{E}_{\mathrm{cool}}$ & $0.24$ & $0.63$
  \enddata
\end{deluxetable}

\begin{figure}
  \centering
  \includegraphics[width=3.0in, keepaspectratio]
  {\figdir/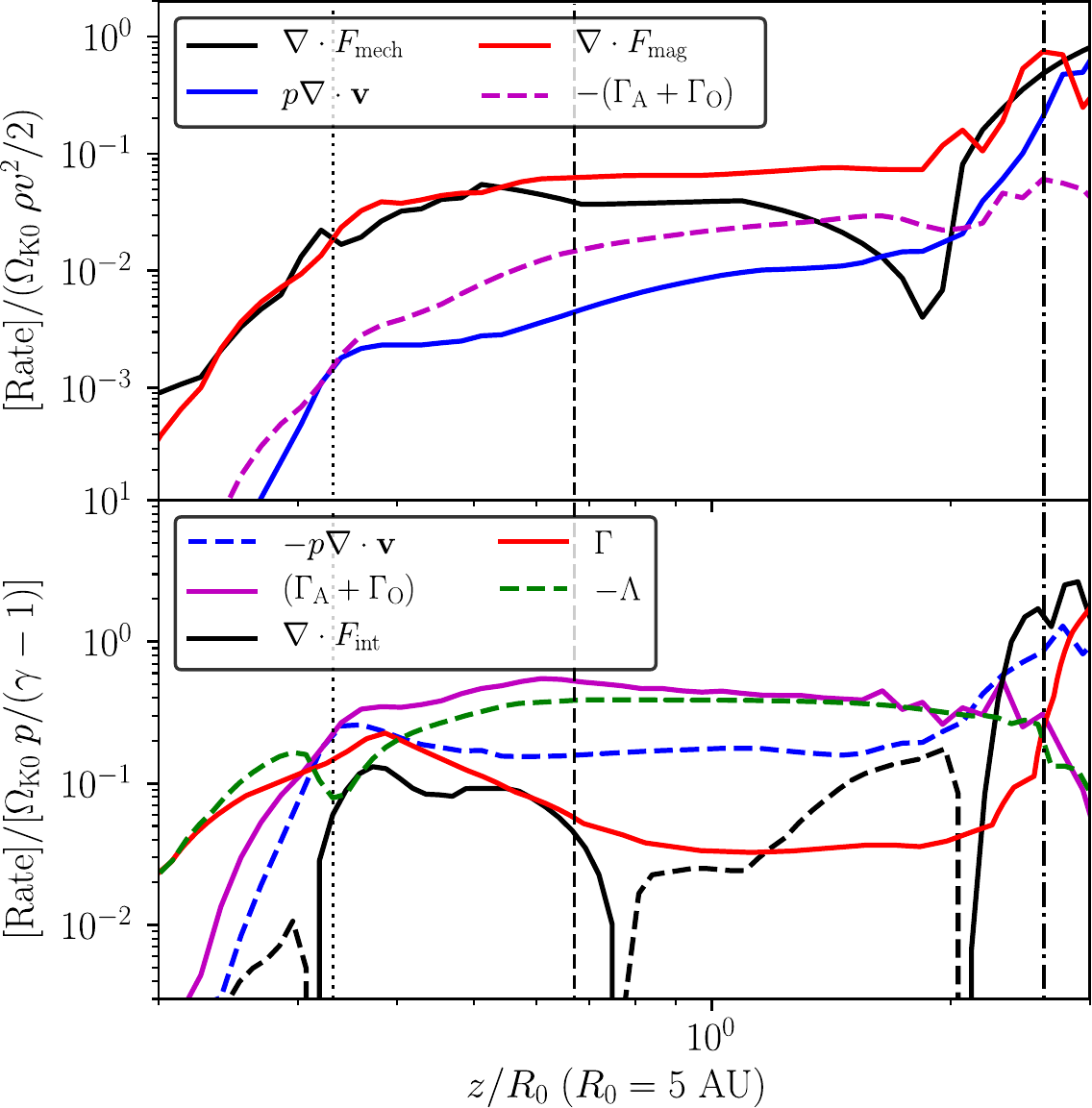}
  \caption{Like Figure~\ref{fig:fiducial_energy} but for
    Model 6 and the field line anchored at $R_0 =
    5~\au$. Vertical dash-dotted lines indicate the
    EUV front.}
  \label{fig:euv1x_energy}
\end{figure}

The energy budget of Model 6 is presented by
Table~\ref{table:euv-ene-budget} and
Figure~\ref{fig:euv1x_energy}, where the global wind budgets
are calculated for the magneto-thermal and EUV winds
separately (we use the EUV front to distinguish the hot EUV
wind from the warm magneto-thermal wind). The
magneto-thermal wind region is similar to but less energetic
than Model 0. The EUV wind region, on the other hand, shows
qualitative differences. Radiative heating overwhelms
magnetic stress and toroidal kinetic power in adding energy
to the wind. As the EUV wind is rather fast and launched
from relatively high altitudes, the majority of its
mechanical power resides in the poloidal motion, instead of
doing work against stellar gravity. 

Table~\ref{table:euv-mass-loss-sp}, compared to its
counterpart for Model 0
(Table~\ref{table:fid-mass-loss-sp}), clearly illustrates
that the EUV-dominated layer contributes a predominantly
ionized flux to the total outflow.  Inspecting
Figure~\ref{fig:euv1x_phase_sp}, one sees that the
distribution functions in the $\{\log_{10}T\}\times\{v_\p\}$
plane of CO and \chem{H_2O}/OH are similar to those of the
fiducial model at the lower-left (low temperature, low
$v_\p$).  The warm \chem{H_2O}/\chem{OH} layer is found at
lower altitudes than in the fiducial model; the conditions
for re-formation of these molecules (see also
\S\ref{sec:fid-wind}) are satisfied in a thin layer dividing
the magneto-thermal wind from the hot EUV wind. EUV
ionization eliminates these molecules in the hot wind, where
the temperature reaches $T\gtrsim 10^{4}~\K$ and the
poloidal velocity $v_\p\gtrsim 15~\kms$. Atomic species
feature much higher velocities and temperatures than the
molecules.  Figure~\ref{fig:euv1x_phase_sp} shows that
relatively significant amounts of neutral atomic O and C
exist within the EUV hot wind layer and extend to similar
temperatures and velocities as the ionized species.  These
relatively high-velocity tails in atomic species may be
useful observationally to distinguish winds with EUV-driven
components from purely magneto-thermal winds.

\begin{deluxetable}{lll}
  \tablecolumns{3} 
  \tabletypesize{\scriptsize}
  \setlength{\tabcolsep}{20pt}  
  \tablecaption{Like Table~\ref{table:fid-mass-loss-sp} but
    for Model 6.
    \label{table:euv-mass-loss-sp}
  }
  \tablehead{
    \colhead{Species} &
    \colhead{$\dot{M}$} &
    \colhead{$\dot{M}/\dot{M}_\wind$} 
    \\
    \colhead{} &
    \colhead{($M_\odot~\yr^{-1}$)} &
    \colhead{}
  }
  \startdata
  \chem{H_2} & $1.07 \times 10^{-8}$ & $6.11 \times 10^{-1}$
  \\ 
  \chem{H} & $5.95 \times 10^{-10}$ & $3.41 \times 10^{-2}$
  \\ 
  \pos{H} & $1.37 \times 10^{-9}$ & $7.83 \times 10^{-2}$ \\
  \chem{He} & $4.46 \times 10^{-9}$ & $2.56 \times 10^{-1}$
  \\ 
  \pos{He} & $5.48 \times 10^{-10}$ & $3.15 \times 10^{-2}$
  \\ 
  \chem{H_2O} & $6.15\times 10^{-15}$ & $3.53 \times
  10^{-7}$ \\ 
  \chem{OH} & $5.59 \times 10^{-14}$ & $3.21 \times 10^{-6}$
  \\ 
  \chem{CO} & $2.84 \times 10^{-11}$ & $1.63 \times 10^{-3}$
  \\ 
  \chem{O} & $4.14 \times 10^{-11}$ & $2.37 \times 10^{-3}$
  \\ 
  \pos{O} & $6.75 \times 10^{-12}$ & $3.87 \times 10^{-4}$ \\
  \chem{C} & $6.30 \times 10^{-12}$ & $3.62 \times 10^{-4}$
  \\ 
  \pos{C} & $2.55 \times 10^{-12}$ & $1.46 \times 10^{-4}$ \\
  \chem{S} & $8.41 \times 10^{-12}$ & $4.83 \times 10^{-4}$
  \\ 
  \pos{S} & $2.82 \times 10^{-12}$ & $1.62 \times 10^{-4}$ \\
  \enddata
\end{deluxetable}

\subsection{Disk Magnetization}
\label{sec:pars-beta}

Model 7 is distinguished from the fiducial model by stronger
magnetization, i.e. ten times lower $\beta_0$.  The
accretion rate is proportionately higher, but
$\dot M_{\rm wind}$ is only $\sim 6$ times higher. The
poloidal wind speed is slightly higher
($\sim 6.7~\km~\s^{-1}$), and the lever arm is larger,
$R_a/R_\wb\sim 1.5$ at all radii, yielding a smaller
ejection index, $\xi\sim 0.5$. The mass density of the wind
is larger, but not sufficient to prevent FUV and X-ray from
reaching deeper than the Alfv\'enic surface. In contrast,
B17 found the FUV front above the Alfv\'enic surface when
$\beta_0$ increases to $10^4$. Toroidal motion in the wind
is super-Keplerian in most places, but the net intertial
(centrifugal plus gravitational) force projected onto field
lines is still negative.

\subsection{MHD Instability: the role of dust}
\label{sec:pars-res-dust}

Model 8 is rather special: it does not have a well-defined
quasi-steady state; instead, it develops instabilities
rather quickly, and these persist throughout the simulation.

\begin{figure}
  \centering
  \includegraphics[width=2.5in, keepaspectratio]
  {\figdir/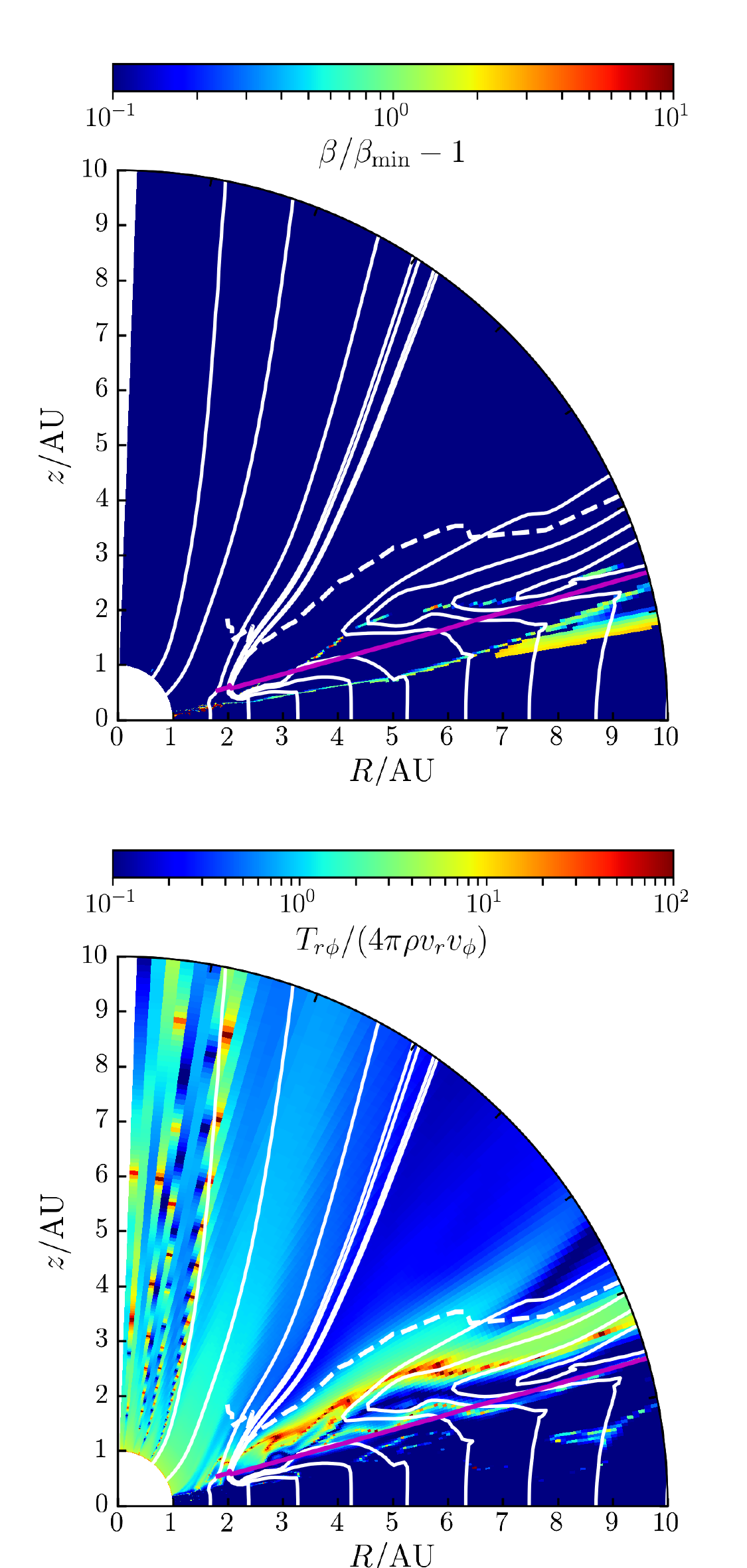}
  \caption{Meridional plots  for
    Model 8  of
    $(\beta/\beta_\min-1)$ (upper panel, see
    eq.~\ref{eq:mri-beta-min}) and
    $|T^{\rm mag}_{r\phi}/(\rho v_rv_\phi)|$ (lower panel)
    showing a snapshot immediately after instabilities
     develop
    ($t\sim 60~\yr$). 
 Streamlines and field
      lines as in Figure~\ref{fig:fiducial_slice}.  }
  \label{fig:dust0.1x_slice}
\end{figure}

Dust grains are crucial in the ionization balance of disks,
especially near the midplane. In general, they assist
recombination by adsorbing charged particles.  If the dust
abundance is reduced, regions with weaker magnetization
become more ionized, which can make the system vulnerable to
MRI. For Model 8, the MRI criterion \eqref{eq:mri-beta-min}
is satisfied in the accretion layer. This is illustrated in
Figure~\ref{fig:dust0.1x_slice}, showing snapshots of
$\beta/\beta_\min$, magnetic stress, and field lines at
$t=60\,yr$, when the instability is active.  Thanks to
increased $\am$, field lines in the accretion layer are
better coupled to inward-moving gas, forming ``spikes''.
Within the unstable layer,
$|T_{r\phi}^{\rm mag}/(\rho v_rv_\phi)|\gtrsim 10$ (see also
Figure~\ref{fig:dust0.1x_slice}). In contrast, for Model 0,
this ratio is always smaller than unity. As axisymmetric
2.5-D simulations cannot deal with saturation of MRI
properly, however, we choose not to over-interpret these
results, and simply mark Model 8 ``unstable''.

\section{Discussions}
\label{sec:discussions}

\subsection{Comparison with B17}
\label{sec:comp-b17}

Although similar to B17 in many respects, by design, this
work features the introduction of real-time consistent
calculations of radiation and thermochemistry. How do these
improvements modify the physical picture of the
magneto-thermal wind mechanism?

The integrated mass-loss rate within the radial range
$1<R/\au <20$ is $2.1\times 10^{-8}~M_\odot~\yr^{-1}$ versus
B17's $7.5\times 10^{-8}M_\odot~\yr^{-1}$.  Yet the
accretion rate throughout the same radial range is lower by
no more than a factor of $1.5$.  Evidently, our winds have
somewhat higher specific angular momentum (larger magnetic
lever arm, smaller ejection index).  The difference should
mainly be attributed to our more consistent temperature and
ionization structures.  Near the wind base, the gas
temperature and $\am$ rise more sharply within the
prescription of B17 than in this work (see
Figures~\ref{fig:fiducial_vert}, \ref{fig:compare_am_mag}).
Shallower gradients can be expected to reduce
$\dot M_{\rm wind}$ because the magnetic and thermal
pressures reach the wind-launching threshold at somewhat
lower gas density.



To better understand the difference in mass-loss rates and
ejection indices between the present models and those of
B17, we have compared their physical properties in more
detail. Figure~\ref{fig:fiducial_force} shows that the the
acceleration by gas pressure ($f_\mathrm{gas}$) is greater
than magnetic pressure ($f_\mathrm{mag}$) at the wind base,
and $f_\mathrm{gas}$ remains comparable to $f_\mathrm{mag}$
until the wind becomes super-Alfv\'enic. B17, in contrast,
had $f_\mathrm{gas}$ making a much smaller or even negative
contribution to wind launching.  Another interesting
comparison involves our Model 6 and the corresponding model
with $\beta_0\simeq 10^4$ in B17 (namely model B40).  The
latter features an FUV front above the Alfv\'enic surface;
FUV photons are completely shielded from the acceleration
region. Our Model 7, however, has FUV photons penetrating
well below the Alfv\'en surface. Those deep-penetrating FUV
photons are partially the reason that Model 6 has greater
$\dot{M}_\wind$ than Model 0 by a factor of 6, while in B17
the corresponding factor is only 3. This difference
emphasizes the value of consistent calculation of
microphysics in non-ideal MHD simulations.

\subsection{Comparison with WG17}
\label{sec:comp-wg17}

Comparisons with WG17 are complicated by the different disk
profile used here (and by B17). Nevertheless, it is
significant that even before the poloidal magnetic field is
applied, Model 6 exhibits
$\dot{M}_\wind \simeq 2.2\times 10^{-9}~M_\odot~\yr^{-1}$
integrated over the range $1 < (r/\au) < 50$ radial range,
thanks to EUV radiation. This $\dot{M}_\wind$ is comparable
to WG17's fiducial result integrated over the range
$2 < (r/\au) < 100$.

Even after the field is applied and Model~5 reaches its
magnetized steady state, there are major similarities to the
unmagnetized models of WG17.  In both cases, the vertical
structure can be described as having three layers: (1) a
relatively cold midplane that is shielded from high-energy
radiation; (2) a warm ($T\sim 10^2-10^3~\K$), largely
molecular ``intermediate layer''; and (3) a hot
($T\sim 10^4~\K$), fast ($v_\p\gtrsim 40~\km~\s^{-1}$)
EUV-dominated wind. In fact, if we measure the mass flux
only in the fast wind, we obtain a result
($2.6\times 10^{-9}~M_\odot~\yr^{-1}$) very similar to that
of the purely hydrodynamic simulations in WG17
($2.5\times 10^{-9}~M_\odot~\yr^{-1}$). The flow in the
EUV-dominated wind is also similar to that in WG17, because
gas pressure is comparable to magnetic pressure
($\beta\sim 10$; see Figure~\ref{fig:euv1x_slice} and
\ref{fig:euv1x_force}).

Nevertheless, including MHD drives the intermediate layer
outwards via the magneto-thermal mechanism, in contrast to
the radially static intermediate layers found by WG17.  The
mass-loss rate in the intermediate layer is approximately
$1.6\times 10^{-8}~M_\odot~\yr^{-1}$, and this gas is
predominantly molecular. Therefore, we observe significant
fraction of molecular species in the gas flowing out of the
simulation domain (Table~\ref{table:euv-mass-loss-sp}) in
Model 6, while in WG17 molecules are depleted by
photodissociation and photoionization before ever reaching
the boundaries.  Molecules are also found at higher
velocities in these models than in those of WG17. For
example, the distribution function of \chem{H_2O}/OH
molecules extends to $\sim 15~\kms$ (see
Figure~\ref{fig:euv1x_phase_sp}), compared to $\sim 10~\kms$
in WG17's figure~4.  In addition, as toroidal fields provide
more pressure, the outflowing ``intermediate layer'' is more
inflated than that of WG17, which gives more the solid angle
to intercept EUV photons and allows the fast wind to start
with less depth of stellar gravitational potential. As a
result, the EUV outflow has slightly higher velocity and
density than WG17.

\subsection{Implications for Observables}
\label{sec:observables}

A major advantage of concurrently evolving a thermochemical
network is that molecular/atomic signals are directly
connected to the kinematics and dynamics of disk outflows,
which holds out the promise that the underlying physics of
PPD dispersal may be constrained by comparing these models
with observations.

Another important possible goal is to distinguish PPD
outflows driven by MHD effects from those driven solely by
photoevaporation. WG17's photoevaporative winds yield
neutral atoms and molecules whose poloidal motion traces
only the thin layer near the EUV wind bases base. These
molecules dissociate before moving a significant distance
within this layer, hence the toroidal motion of those
molecules is almost Keplerian.  Magneto-thermal winds, on
the other hand, are substantially neutral and molecular. At
high altitudes ($\theta\lesssim 0.6$), the neutral molecular
winds deviate significantly and systematically from
Keplerian. With ALMA data, \citet{2013A&A...555A..73K,
  2014ApJ...792...68S, 2016Natur.540..406B} have already
found significant large-scale molecular outflows. Optical
and infrared observations on forbidden lines of atoms/ions
(e.g. [\ion{O}{I}] $\lambda 6300$) are also intepreted as
signatures of PPD winds (e.g. \citealt{1995ApJ...452..736H,
  2014A&A...569A...5N, 2016ApJ...831..169S,
  2018arXiv181003366F}; Banzatti et al. 2018, submitted).
In addition, \citet{2018ApJ...856..117F} constrained the
amplitude of turbulence inside the disk of TW Hya and
concluded that $\alpha<0.007$. These result may constrain
the wind-launching and accretion mechanisms by comparing
with models, such as ours, that predict the kinematics and
dynamics of the observed species.

\section{Summary}
\label{sec:summary}

We have studied PPD outflows by conducting axisymmetric
global simulations that combine non-ideal MHD (excepting the
Hall effect) with consistent ray-tracing radiative transfer
and thermochemistry.  Our fiducial model, whose parameters
broadly follow B17, features a warm
($T\sim 10^2$-$10^3~\K$), predominantly molecular,
magneto-thermal wind, driven by the gradient of toroidal
magnetic pressure Such wind has mostly sub-Keplerian
toroidal motion, with typical radial velocity
$\lesssim 5\,\kms$ and a mass-loss rate comparable to the
magnetically driven accretion rate.  By varying luminosities
of high energy photons, especially soft-FUV and Lyman-Werner
photons, we find that the wind mass loss is affected by the
ionization fraction and ambipolar parameter near the wind
base, though, overall rates of mass-loss and accretion are
relatively robust against variations in luminosities. Adding
EUV photons adds a fast ($v_\p\gtrsim 50~\km~\s^{-1}$), hot
($T\gtrsim 10^4~\K$), ionized component to the wind at high
latitudes.  For plausible EUV luminosities, the pressure of
the fast atomic wind compresses the latitudinal range of the
molecular wind and reduces its mass-loss rate by about a
half.  Reducing the dust abundance may raise the ionization
fraction near the wind base, leading to MRI-like
instabilities.  This illustrates that the structure and
dynamics of magneto-thermal winds may well depend
sensitively on subtle details of thermochemistry.

In the future we expect to explore several other aspects of
problems as natural extensions to this work. Adding Hall
effect will likely substantially change the flow structure
and magnetic flux transport in the disk, affecting wind
launching (\citealt{2017ApJ...836...46B}, B17). Because B17
found that the Hall effect tends to break the reflection
symmetry about the midplane, this in turn may require better
treatment of midplane thermochemistry.  Adsorption and
desorption of volatile species on surfaces of dust grains,
processes that we have so far neglected, may change the
thermochemical conditions at the wind base, and in turn
affect the wind dynamics.  Monte-Carlo modeling of
scattered/re-emitted photons (especially infrared and X-ray
photons), instead of simplified recipes, may be needed.
Some parts of the parameter space show signs of MRI
instabilities, which should properly be studied by
3-dimensional global simulations. In fact, even the laminar
models may be subject to 3D kink instabilies in regions
where $B_\phi/B_\p\gg 1$.  Modeling the molecular and atomic
lines implied by these simulations will help us to better
constrain our models by comparison to observations and
improve our understanding of the dispersal and accretion of
PPDs.

\vspace*{20pt} This work was supported by NASA under grant
17-ATP17-0094/80NSSC18K0567 and by Princeton University's
Department of Astrophysical Sciences.  We thank our
colleagues: Phil Armitage, Bruce Draine, Eve Ostriker, James
Stone, and Kengo Tomida, for discussions and for detailed
comments on preliminary versions of this paper. We also
thank Oliver Gressel who inspired the explorations in
non-ideal MHD heating in this work.

\appendix

\section{Additional Details of Radiation and
  Thermochemical Processes}
\label{sec:appx-chem}

This appendix provides details concerning the treatment of
some thermochemical and radiation processes, especially
treatments that have been improved since WG17.

\subsection{Ionization by Diffuse High Energy Photons and
  Radioactive Decay}
\label{sec:appx-diff-ionize}

We follow the recipes in \citet{Bai+Goodman2009}. The
effective ionization rate by diffuse sources of ionization
is
\begin{equation}
  \label{eq:appx-diff-ionize}
  \xi^\eff = \xi^\eff_\mathrm{X-ray} + \xi^\eff_\mathrm{CR} +
  \xi^\eff_\mathrm{SLR}\ .
\end{equation}
The term $\xi_\mathrm{X-ray}$ represents ionization by
down-scattered X-ray photons and is approximated by a
fitting formula based on \citet{Igea+Glassgold1999},
\begin{equation}
  \label{eq:appx}
  \xi^\eff_\mathrm{X-ray} = \left( \dfrac{R}{\au} \right)^{-2.2}
  \left(\dfrac{L_\mathrm{X}}{10^{29}~\erg~\s^{-1}}\right)^2
  \left\{
    \xi_1 \left[ \e^{-(N_{\H1}/N_1)^\alpha}
      + \e^{-(N_{\H2}/N_1)^\alpha} \right]
    + \xi_2 \left[ \e^{-(N_{\H1}/N_2)^\beta}
      + \e^{-(N_{\H2}/N_2)^\beta} \right]
  \right\}\ .
\end{equation}
Here $L_\mathrm{X}$ is the X-ray luminosity (in our
simulations, the luminosity in the $3~\keV$ energy band).
The terms involving $\xi_1$ and $\xi_2$ describe attenuation
by absorption and scattering, respectively, while $N_{\H1}$
and $N_{\H2}$ are the vertical column densities of hydrogen
nuclei measured above and below the point of interest (thus
accounting for penetration from both sides of the disk).
$\xi_1 = 4.0\times 10^{-12}~\s^{-1}$,
$\xi_2 = 2.0\times 10^{-15}~\s^{-1}$,
$N_1 = 1.5\times 10^{21}~\cm^{-2}$,
$N_2 = 7.0\times 10^{23}~\cm^{-2}$, $\alpha = 0.5$, and
$\beta = 0.65$.  For the sake of simplicity we approximate
the vertical density profile as Gaussian and take,
\begin{equation}
  \label{eq:appx-nh-estimation}
  \begin{split}
  & N_{\H\,1}(z) = n_\H(z) \int_z^{+\infty}\d z'\
  \e^{(z^2-z'^2)/2h^2} \sim h n_\H(z)\ ,
  \\
  & N_{\H\,2}(z) =  n_\H(z) \int_{-\infty}^{+\infty}\d z'\ 
  \e^{(z^2-z'^2)/2h^2} - N_{\H\,1}(z)
  \sim 2.5 h~\e^{z^2/2h^2} n_\H(z) - N_{\H\,1}(z)\ ,
  \end{split}  
\end{equation}
where
$h$$\equiv\cs/\Omega$ is the local disk scale height based
on the midplane isothermal sound speed $\cs$, and
$n_\H$ is the local number density of hydrogen nuclei. Rough
estimates as eqs. \eqref{eq:appx-nh-estimation} are, we
verify for each model that such estimates are different from
exact results by no more than $\sim
20~\%$ near and below the wind base.

The cosmic ray term,
$\xi^\eff_\mathrm{CR}$, is approximated using the updated
recipes in \citet{1981PASJ...33..617U}
\begin{equation}
  \label{eq:appx-cr-ionize}
  \xi^\eff_\mathrm{CR} = 10^{-17}~\s^{-1}\times
  \left( \e^{-N_{\H1}/N_\mathrm{CR}} +
  \e^{-N_{\H2}/N_\mathrm{CR}} \right)\ ,
\end{equation}
where $N_\mathrm{CR} = 5.7\times
10^{25}~\cm^{-2}$.  The contribution of short-lived
radioactive nuclei is presumed constant,
$\xi^\eff_\mathrm{SLR} = 6.0\times
10^{-19}~\s^{-1}$. The effective ionization rate is divided
among three channels according to
\begin{equation}
  \label{eq:appx-ionizing-channel}
  \begin{split}
    \chem{H_2} \rightarrow \H + \pos{\H} +
    e^-\ ,\quad & \xi = \xi^\eff\ ;
    \\
    \H \rightarrow \pos{\H} + e^-\ ,\quad &
    \xi = 0.50~\xi^\eff\ ;
    \\
    \chem{He} \rightarrow \pos{\chem{He}} + e^-\ ,\quad &
    \xi = 0.84~\xi^\eff\ ;
  \end{split}
\end{equation}
Admittedly these are rather rough estimates. However, the
focus of this paper is on the disk atmosphere and wind zone
, where the X-rays are scarcely attenuated and the other
sources of ionization above---cosmic rays and
radioactivities---are less important.

\subsection{Self-/Cross-Shielding and Absorption of FUV
  Photons}
\label{sec:appx-shielding}

Some photoreactions, especially photoionizations and
photodissociations by absorption lines (e.g.  in the LW
band), are subject to self-/cross-shielding effects. For the
dominant species in the intermediate zone, \chem{H_2}, we
adopt the scheme in WG17, which treats the photo-pumping
$\chem{H_2} + h\nu(\mathrm{LW})\rightarrow \chem{H_2}^*$
with the self-shielding recipes in
\citet{1996ApJ...468..269D}. Such photo-pumping is the
starting point of some important subsequent processes,
including collisional heating, dissociation, and spontaneous
decay. For other species, we adopt the tabulated
self-shielding and cross-shielding coefficients (by
\chem{H_2} and by C if applicable) obtained by
\citet{2017A&A...602A.105H} for the photoreactions of the
following species: C, CO, S, Si, and SiO.

\subsection{Pre-absorption of Radiation at the Inner
  Radial Boundary}
\label{sec:appx-pre-absorb}

As our simulations exclude the central zone, ionizing
radiation is not attenuated by any outflow or raised disk
rim at $r < r_\mathrm{in}$. At low latitudes in the
predominantly neutral outflow near the inner radial
boundary, this absence of attenuation could cause
unphysically high ionization fractions, which compromise the
credibility and even stability of our simulations. We
therefore use a ``pre-absorption'' scheme to account for
such attenuation: for each energy bin, the absorption
optical depth of a ray at co-latitude $\theta$ from the
central source to the inner boundary $r_\mathrm{in}$,
$\tau_\mathrm{in}(h\nu,\theta)$, is estimated as follows. We
assume that the ratio
$\tau_\mathrm{out}(h\nu,\theta)/\tau_\mathrm{in}(h\nu,\theta)$
equals $\zeta^{-1} r_\mathrm{in} / r_\mathrm{out}$, where
$\zeta$ is a constant factor, and
($\tau_\mathrm{out}(h\nu,\theta)$ is the absorption optical
depth between $r_\mathrm{in}$ and $r_\mathrm{out}$ along the
same ray.  Hence,
\begin{equation}
  \label{eq:pre-abs}
  \tau_\mathrm{in}(h\nu, \theta) \simeq
  \zeta \dfrac{r_\mathrm{in}}{r_\mathrm{out}}
  \tau_\mathrm{out}(h\nu,\theta)\ ;\quad 
  \tau_\mathrm{out}(h\nu,\theta) = 
  \ln \left[
    \dfrac{F_\mathrm{in}(h\nu, \theta)}
    {F_\mathrm{out}(h\nu, \theta)} 
  \right]\ ,
\end{equation}
where $F_\mathrm{in}(h\nu,\theta)$ and
$F_\mathrm{out}(h\nu,\theta)$ are the photon fluxes along
the ray at the inner and outer boundaries,
respectively. $F_\mathrm{in}$ and $F_\mathrm{out}$ are
measured from the previous time step.  We tested several
values of $\zeta$ from $10^{-1}$ to $10^1$, confirming that
the results of simulations are not sensitive to $\zeta$. For
convenience we choose $\zeta = 1$ throughout this work.

\bibliography{mhd_wind}
\bibliographystyle{apj}

%
\end{document}